%% file: main.tex
\documentclass[twocolumn, times, tighten, trackchanges]{aastex631}
\pdfoutput=1
\usepackage{amsmath}

\usepackage{savesym}
\savesymbol{tablenum}
\usepackage[separate-uncertainty=true]{siunitx}
\restoresymbol{SIX}{tablenum}   

\usepackage{natbib}
\newcommand{\nh}{\ensuremath{N(\text{H})}}
\newcommand{\nhi}{\ensuremath{N(\text{\ion{H}{1}})}}
\newcommand{\nhtwo}{\ensuremath{N(\text{H}_2)}}
\newcommand{\denh}{\ensuremath{n(\text{H})}}

\newcommand{\ebv}{\ensuremath{E(B-V)}}
\newcommand{\av}{\ensuremath{A(V)}}
\newcommand{\ak}{\ensuremath{A(1000)}}
\newcommand{\akav}{\ensuremath{A(1000)/A(V)}}
\newcommand{\abumpav}{\ensuremath{A(2175)/A(V)}}
\newcommand{\nhav}{\ensuremath{N(\text{H}) / A(V)}}
\newcommand{\rv}{\ensuremath{R(V)}}
\newcommand{\rvi}{\ensuremath{1/R(V)}}
\newcommand{\fhtwo}{\ensuremath{f(\text{H}_2)}}
\newcommand{\cav}[1]{\ensuremath{c^{A(V)}_{#1}}}

\newcommand{\lya}{{Ly$\alpha$}}

\newcommand{\hi}{\ion{H}{1}}

\DeclareSIUnit\mag{mag}
\DeclareSIUnit\angstrom{\text{\AA}}
\DeclareSIUnit\erg{\text{erg}}

\defcitealias{2009ApJ...705.1320G}{GCC09}

\begin{document}

\title{Far-ultraviolet Dust Extinction and Molecular Hydrogen in the Diffuse Milky Way Interstellar Medium}

\correspondingauthor{Dries Van De Putte}
\email{dvandeputte@stsci.edu}

\author[0000-0002-5895-8268]{Dries Van De Putte}
\affiliation{Space Telescope Science Institute, 3700 San Martin
  Drive, Baltimore, MD, 21218, USA}

\author{Stefan I.\ B.\ Cartledge}
\affiliation{Dept.\ of Physical Sciences, MacEwan
  University, 10700 - 104 Ave., Edmonton, AB T5J 4S2}

\author[0000-0001-5340-6774]{Karl D.\ Gordon}
\affiliation{Space Telescope Science Institute, 3700 San Martin
  Drive, Baltimore, MD, 21218, USA}
\affiliation{Sterrenkundig Observatorium, Universiteit Gent,
  Gent, Belgium}

\author[0000-0002-0141-7436]{Geoffrey C.\ Clayton}
\affiliation{Department of Physics \& Astronomy, Louisiana State University,
  Baton Rouge, LA 70803, USA}

\author[0000-0001-6326-7069]{Julia Roman-Duval}
\affiliation{Space Telescope Science Institute, 3700 San Martin
  Drive, Baltimore, MD, 21218, USA}
  
\defcitealias{GCC09}{2009ApJ...705.1320G}

\begin{abstract}

We aim to compare variations in the full-UV dust extinction curve (912-3000 \AA), with the H\ I/H$_2$/total H content along diffuse Milky Way sightlines, to investigate possible connections between ISM conditions and dust properties.
We combine an existing sample of 75 UV extinction curves based on IUE and FUSE data, with atomic and molecular column densities measured through UV absorption.
The H$_2$ column density data are based on existing Lyman-Werner absorption band models from earlier work on the extinction curves.
Literature values for the H\ I column density were compiled, and improved for 23 stars by fitting a Ly$\alpha$ profile to archived spectra.
We discover a {strong correlation} between the H$_2$ column and the far-UV extinction,
{and the underlying cause is a linear relationship between H$_2$ and the strength of the far-UV rise feature.}
This extinction does not scale with H\ I, and the total H column scales best with \av\ instead.
{The carrier of the far-UV rise} therefore coincides with molecular gas, and further connections are shown by comparing the UV extinction features to the molecular fraction.
Variations in the gas-to-extinction ratio \nhav\ correlate with the UV-to-optical extinction ratio, and we speculate this could be due to coagulation or shattering effects.
Based on the H$_2$ temperature, the strongest far-UV rise strengths are found to appear in colder and denser sightlines.
\end{abstract}


\section{Introduction}

The presence of dust grains in the interstellar medium (ISM) has several effects, through which the the state of the medium is influenced. 
Dust provides provides shielding by absorbing the UV photons in the Lyman and Werner bands (912 - \SI{1100}{\angstrom}), which dissociate H$_2$ \citep{2014ApJ...790...10S}.
If there are more small grains that absorb strongly in the far-UV, then this shielding will be more efficient per dust mass unit.
Dust also facilitates the formation of H$_2$, on the surfaces of grains \citep{1971ApJ...163..155H, 2017MolAs...9....1W}.
The H$_2$ formation rate is proportional to the available dust surface area, which is why this process is more efficient when a larger fraction of the dust content consists of smaller grains, because of their higher surface area-to-volume ratio \citep{1983ApJ...269L..57S}.
Because both the shielding and the formation efficiency of H$_2$ are enhanced by the presence of small grains, and the extinction curve of dust in the UV depends on the grain size distribution, there could be an observable connection between UV dust extinction and the H I/H$_2$ balance of the gas.

Extinction curves are the key observable providing information about the optical properties and size distribution of dust.
Processes such as shattering, coagulation, and accretion can change the size distribution of dust grains, which in turn results in a variety of extinction curve shapes, as demonstrated in dust evolution models of galaxies \citep{2009A&A...502..845O, 2012MNRAS.422.1263H, 2014MNRAS.440..134A, 2019MNRAS.482.2555H, 2020MNRAS.491.3844A}.
One of the key questions in the study of dust extinction curves, is the origin of two prominent UV features: a broad bump at \SI{2175}{\angstrom} in the near-UV (NUV bump), and an steepening of the curve at far-UV wavelengths (FUV rise) which continues all the way to \SI{912}{\angstrom} \citep{1986ApJ...307..286F, 1988ApJ...328..734F}.
The strength and rate of occurrence of these features has been observed to be different between the Milky Way (MW), Large Magellanic Cloud (LMC), and Small Magellanic Cloud (SMC) \citep{2003ApJ...594..279G}.

When the average grain size increases, the extinction curve becomes flatter towards shorter wavelengths.
A commonly used quantity to measure the slope of the extinction curve, and make conclusions about changes in the average grain size, is the total-to-selective extinction ratio $\rv = \av / \ebv$.
The relationship between characteristics of the size distribution of dust models and the corresponding $\rv$ is demonstrated in e.g.\ \citet{2018A&A...611A...5S}.
Observations in the mid-IR of individual molecular clouds \citep[e.g.][]{2009ApJ...690..496C, 2010A&A...511A...9S} find that $\rv$ is greater in the inner regions with high extinction, suggesting that larger grains are formed more efficiently in the colder and denser parts of clouds.
UV extinction curve observations of the Taurus molecular cloud TMC-1 show that the \SI{2175}{\angstrom} extinction bump is prominent in diffuse regions, but becomes minimal inside the dense clump \citep{2004ApJ...602..291W}.
The sightlines studied in this work are diffuse, with $\av < 3$, which means that are are not observing deep into such clouds, but rather the edges, where photodissociation regions occur (PDRs, \citealt{1999RvMP...71..173H}).
In earlier work \citep{2020ApJ...888...22V}, we measured the extinction in different locations of the IC\,63 PDR using broadband photometry of its background stars, and found that $\rv$ is larger near the illuminated edge.
It is unclear if the observed $\rv$ variations in this PDR originate from accretion processes similar to those assumed for dense molecular clouds.
At larger scales, for a combined sample LMC and SMC sightlines, correlations between the gas-to-extinction ratio and the steepness of the curve were found \citep[Figure 9]{2003ApJ...594..279G}.
In this work we aim to examine if similar dust evolution effects occur in the diffuse Milky Way ISM, by investigating \rv\ and other extinction curve shape parameters.

{In addition to tracing the grain size with \rv, the extinction curve also contains information about the grain composition.
While it is still unclear what the exact carriers of the UV bump and rise features are, most models assume that they are associated with carbonaceous grains, as opposed to silicate grains.
For example, in recent work by \citet{2021ApJS..257...63Z}, models based on \cite{2001ApJ...548..296W} are fitted to a sample of extinction curves that partially overlaps with the one used in this work.
The silicate component of their model has a flatter extinction curve, while the carbonaceous component has the UV features.
Under those assumptions, the strength of the bump and rise determine the amount of carbon in the dust.
In the SMC on the other hand, the bump is generally weak, while the FUV rise is still strong. 
Depletion measurements have shown that some SMC dust has very little Si \citep{2015ApJ...811...78T}, and that the relative amount of dust consisting of Polycyclic Aromatic Hydrocarbons (PAHs) is significantly lower than the MW \citep{2017ApJ...838...85J}. 
This indicates that the grains causing the bump and rise exhibit different behaviors depending on the environment, despite the fact that both features are thought to be carbonaceous in nature.
A more subtle clue about the nature of the grains is the bump width, which has been found to vary depending on the environment, being narrower in the diffuse medium and broader in dark clouds.
These changes are typically attributed to coatings forming on the grains \citep{1991AJ....101.1021C, 2004ApJ...616..912V}.
By comparing the UV extinction features and the H$_2$ content of the gas, we aim to learn about the relationship between carbonaceous grains and molecular gas.}

A secondary goal, is to search for direct evidence for dust growth in the ISM, a process which is found necessary to explain the evolution of the dust content of galaxies \citep{2008A&A...479..453Z, 2009ASPC..414..453D, 2012ApJ...748...40B, 2014MNRAS.441.1040R, 2016ApJ...831..147Z}.
The discrepancy between the observed amount of dust, and the dust production and destruction rates, can be solved if dust grains grow in mass and size by accreting gas-phase metals onto their surfaces \citep{2016MNRAS.457.3775M}.
Direct evidence for dust growth, i.e.\ atoms moving from the gas to the solid phase, has been found via depletion measurements.
This type of measurement consists of determining which fraction of each metal is locked up in dust grains, based on measurements of the gas-phase abundances of those metals \citep{1996ARA&A..34..279S}.
The degree of depletion across different metals is mostly tied to a single parameter called $F_*$ \citep{2009ApJ...700.1299J}, and this parameter correlates well with the average number density along the sightline \denh, but less so with the molecular fraction \fhtwo.
The gas-to-dust ratio is also {a} tracer for this type of growth, and this metric has been found to be different between the diffuse and the dense ISM in the Magellanic Clouds \citep{2014ApJ...797...86R,2017ApJ...841...72R,2021ApJ...910...95R}, providing evidence that grain growth occurs mostly in high-density regions of galaxies.

The variations in the extinction likely influence the way the ISM is affected by a UV radiation field.
In numerical models of PDRs the depth of the \hi-H$_2$ transition is commonly expressed in terms of the optical depth \av\ \citep{2007A&A...467..187R}.
However, due to changes in the grain size distribution, the UV-to-optical extinction ratio can differ between sightlines, between PDRs, or even between different locations in the same PDR.
Figure 3 of \citet{1990ARA&A..28...37M} shows how the penetration of UV radiation changes for different $\rv$, at constant $\av$.
Since UV radiation and its attenuation by dust and gas are the main drivers of the physics in a PDR, the outcome can be very different for clouds with similar \av\ but different dust populations.
Radiative transfer simulations of PDRs show that a spatially varying extinction law can produce substantially different locations of the \hi-H$_2$ transition \citep{2007A&A...467....1G}.

Coming back to the main goal of this work, we aim to study how interstellar dust influences and responds to the different environments in the diffuse interstellar medium of the Milky Way, with a focus on H$_2$.
To probe the relationship between details of the UV extinction curves, the H$_2$ content, and the gas-to-dust ratio, we need a sample of sightlines with measurements of the necessary quantities: the extinction at every UV wavelength $A(\lambda)$, and the column densities \nhi\ and \nhtwo.
The sample of 75 diffuse Milky Way sightlines by \citet{2009ApJ...705.1320G}  (hereafter GCC09) is ideal for this.
It is currently the largest sample with extinction curves measured using FUSE far-UV data (912-1200 \AA), and covers a large range of $\rv$ values $\sim 2.6 - 5.5$.
Preliminary work, investigating far-UV extinction properties in the Galaxy and the Magellanic Clouds with IUE data ($\sim$ 1170 - \SI{3200}{\angstrom}), includes 78 Milky Way sightlines by  \cite{1990ApJS...72..163F}, and several LMC sightlines \citep{1999ApJ...515..128M} and SMC sightlines \citep{1998ApJ...500..816G}.
Later, FUSE data were used to extend extinction curves to shorter wavelengths. \citet{2005ApJ...625..167S} studied the far-UV extinction curves along nine Milky Way sightlines and \citet{2005ApJ...630..355C} presented a study of 9 far-UV extinction curves in the Magellanic Clouds.
The GCC09 paper extended the sample of Milky Way sightlines with FUSE data to 75 stars, and used a more mature FUSE calibration, a better correction for the H$_2$ absorption, and a more complete set of comparison stars.
These improvements in data quality have exposed the need for improved extinction curve parameterizations.
In GCC09, the reformalization of the CCM family of curves \citep{1989ApJ...345..245C} was initiated, concentrating on the FUV spectrum of stars observed using FUSE.
In this work, we effectively extend the original analysis.
While GCC09 focused on correlations between extinction curve properties, we will investigate correlations between those properties and quantities related to the gas. 

The number of previous observational studies including both the details of H$_2$ and the extinction curves is rather limited.
A correlation between the molecular fraction and the bump width was found by \cite{2002ApJ...577..221R}, using a sample of 23 sightlines.
This study also found a less significant correlation between the strength of the FUV rise and the molecular fraction.
Our work significantly improves the statistical significance for the latter correlation.
Another study investigated the relation between \rv\ and H$_2$, for 38 sightlines \citep{2009ApJS..180..125R}.
They did not find significant evidence for a general trend where the molecular fraction is lower for high-\rv\ sightlines.
With our sample, the number of sightlines is about doubled, and we examine if this conclusion holds.

In Section \ref{sec:data}, we describe data used for our investigation of the GCC09 sample 75 diffuse Milky Way sightlines. 
This includes the data provided by the GCC09 paper, the H$_2$ absorption models that were fit to the FUSE data to determine \nhtwo, and the \lya\ absorption profile fitting method used to determine \nhi. 
We also describe how several derived quantities were calculated, and how we deal with the uncertainties and correlations between those quantities.
In Section \ref{sec:results}, we show the correlations that were found when comparing the extinction and the gas contents.
The interpretation of these results is discussed in Section \ref{sec:discussion}, and to conclude, the main results and their interpretation are summarized in Section \ref{sec:conclusions}, followed by a brief future outlook.

\section{Data}
\label{sec:data}

\subsection{Sample}
In GCC09, the original paper discussing this sample, it is described in detail how the sightlines were selected. In short they were selected from the set of all Galactic O and B stars observed by FUSE, that also have sufficient observations to provide spectral coverage from the infrared (IR) through the FUV.
This sample includes 75 reddened stars and 18 lightly reddened stars, the latter serving as comparison stars.
The data we collected for these sightlines are summarized in two tables.
Table~\ref{tab:h2dat} contains molecular hydrogen column densities for rotational levels up to $J=7$. 
The level populations result from fitting an H$_2$ absorption model to FUSE spectra, as described in Section \ref{sec:h2models}.
These models were created in the context of the GCC09 paper, to remove H$_2$ absorption features from the far-UV extinction curves.
However, the resulting level populations and a detailed description of the fitting method went unpublished until now.

The total molecular and atomic hydrogen column densities, as well as several derived quantities for the sightlines, are summarized in Table~\ref{tab:gasdetails}.
Sections \ref{sec:hi} and \ref{sec:properties} describe how these data were obtained.

\input{h2_edit.tex}
\clearpage
\input{hi_edit.tex}

\subsection{H$_2$ Modeling}
\label{sec:h2models}

The complete set of FUSE data contains 110 sightlines. In GCC09, 75 of them have extinction curves, 18 were used as comparison stars, and for 9 of them it was not possible to derive extinction curves.
We publish the H$_2$ results for all of them, and have sorted the sightlines accordingly in Table~\ref{tab:h2dat}, but we will not use the last 9 sightlines in the rest of this work.

The H$_2$ column densities in each rotational excitation level, the line width parameter $b$, and the heliocentric velocity of the profile center define the absorption model for each sightline.
In general, an effort was made to keep the models as simple as possible.
In the absence of clear evidence for multiple velocity components in the absorption, a single component is assumed.
Single component models are used for 92 of the 110 sightlines, two component models are used for 14 sightlines, two sightlines are described by three component models, one by a four component model, and one is composed of six absorption components.
To determine the number of components needed, lines of other gas species were used, in data from earlier work on elemental abundances \citep{2006ApJ...641..327C} or information from similar papers such as \citet{2002ApJ...577..221R}.
When other species significantly indicated multiple absorption components, the number is based on the strongest absorbing components.
Otherwise, the modeling started with a single component, and the number of components was increased until the continuum of the lines could be reconstructed sufficiently.

The procedure for constructing these molecular hydrogen models was patterned after a common approach applied to FUSE data \citep{2001ApJ...555..839R,2002ApJ...577..221R} that we have previously used in analyses of FUV extinction curves \citep{2005ApJ...625..167S,2005ApJ...630..355C,2009ApJ...705.1320G}.
In the context of the GCC09 paper, the FUSE observations from the archive were processed using CALFUSE (v3.0).
The GCC09 paper did not provide the details about the reduction and H$_2$ modeling of these FUSE data, and instead referred to a future paper which went unpublished.
We therefore provide the details in this work.

In cases where the different observations or FUSE channels produced conflicting flux levels for a given wavelength (e.g.\ due to channel misalignment), the fluxes for these observations or channels were scaled by a single factor to match the maximum observed flux across the regions of spectral coverage overlap.
Then, for each single channel of each observation, the calibrated spectrum was cross-correlated with either the LiF1A or LiF2B spectrum.
The latter choice applied to the LiF1B and LiF2A spectra, whose wavelength coverage does not overlap with LiF1A.
In this fashion, each spectrum could be shifted such that it conformed to the LiF1A wavelength solution across the LiF1A bandpass, and could then be averaged together across the entire FUSE wavelength window.
Note that the solutions for each segment are technically not compatible with a simple linear shift;
nevertheless, the quality of the data resulting from the averaging process suffices for the purpose of measuring molecular hydrogen column densities.
A single FUSE spectrum was generated for each sightline by averaging together all observations of a given star on a wavelength-by-wavelength basis, weighted by exposure time.

The FUSE spectrum for a given sightline was then processed using a code that allows the user to concurrently identify absorption features due to the various rovibrational H$_2$ transitions ($J' \rightarrow 0$ and $v' \rightarrow 0$) in the FUSE 910--1190{\AA} bandpass.
The equivalent widths were measured assuming a background stellar spectrum fit by a low-order polynomial.
The order of the polynomial varies from sightline to sightline, depending on the stellar continuum, but was at most second order.
For each segment of the H$_2$ fits, there was no concrete evidence for higher-order variations on the scale of the wavelength windows used.
The data for each absorption component along the sightline are listed in Table~\ref{tab:h2dat}, omitting lines that were indistinguishably blended with other absorption features.
For the sightlines where multiple H$_2$ components were needed, the table has multiple rows associated with a single star name.

\begin{figure*}
  \centering
  \plotone{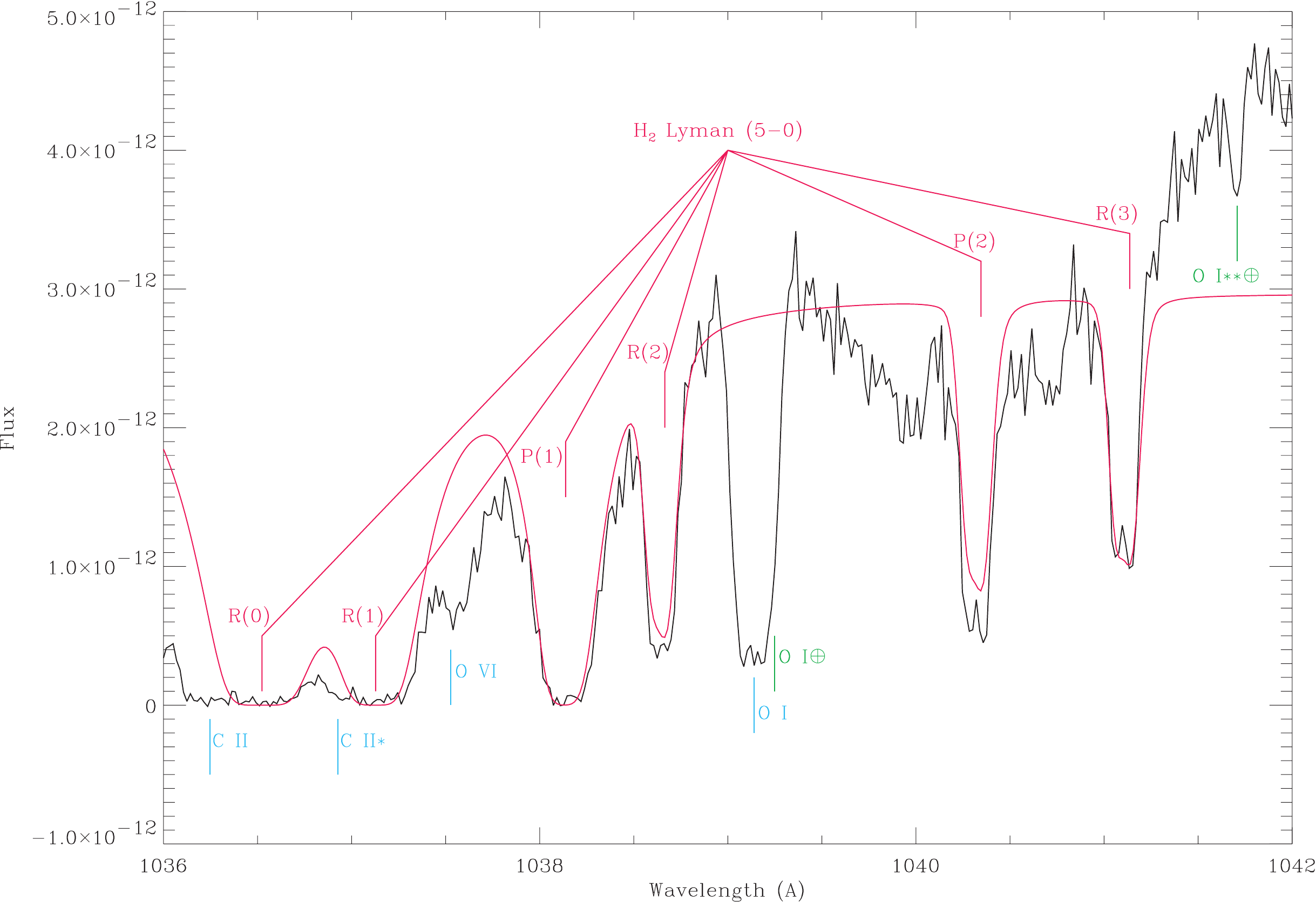}
  \caption{Example of the H$_2$ absorption line modeling.
    The flux units are \si{\erg\per\square\cm\per\s\per\angstrom}.
    The red line shows a transmission model for the Lyman band of H$_2$, which consists of transitions from the electronic ground state $X$, to the excited electronic state $B$ ($2p ^1\Sigma_u^+$).
    This example model includes contributions by H$_2$ molecules with rotational quantum numbers $J = 5 - 0$.
    The H$_2$ lines are labeled with the $J$ of the lower state, and a branch name indicating whether $J$ increases by 1 (R-branch) or decreases by 1 (P-branch) during the electronic transition, as constrained by the selection rules.
    The absorption lines by several atoms and ions have been indicated in blue for the interstellar gas, and in green  with a ``$\bigoplus$'' symbol for lines caused by the earth's upper atmosphere.}
  \label{fig:h2model}
\end{figure*}

In cases where multiple components were evident, particular care was taken to separate the amount of absorption associated with each line-of-sight cloud using the profile-fitting code \textit{fits6p} \citep{1991ApJS...75..425W,1995PASA...12..239M}.
In addition, this code was used on the spectra from every sightline to determine reasonable $b$-values for each absorption component along a sightline using weak $J$=0--0 transitions longward of \SI{1110}{\angstrom}.
The errors on the equivalent widths are also provided by the \textit{fits6p} code, after the determination of the $b$ values, and these include contributions from both the continuum fitting and the noise in the data.
Each table was then subjected to a curve-of-growth (CoG) analysis which generated a $b$-value and $J$=0--7 molecular hydrogen column density solution using the multidimensional IDL fitting routine AMOEBA\footnote{https://www.l3harrisgeospatial.com/docs/amoeba.html}.
Generally, however, the CoG analysis was limited to $J \geq 3$ because the $J \leq 2$ lines were saturated to the extent that they possessed broad damping wings that the equivalent width measurement algorithm could not reliably distinguish from the stellar continuum.
An example spectrum and the resulting H$_2$ model is shown in Figure~\ref{fig:h2model}.
The H$_2$ columns for each of the rotational levels ($J$=0--7) were summed to give the total column density for each sightline, as listed in Table~\ref{tab:gasdetails}.
The table also distinguishes sightlines used as comparison stars in the extinction analysis from the more heavily reddened program stars \citep{2009ApJ...705.1320G}.

\subsection{Atomic Hydrogen}
\label{sec:hi}

Table~\ref{tab:gasdetails} displays the column densities of total (\nh), atomic (\nhi), and molecular hydrogen (\nhtwo).
Wherever it was possible, literature values for \nhi\ were adopted, as indicated by the reference indices in the rightmost column of Table~\ref{tab:gasdetails}.
For 23 of the sightlines, the available sources only listed \nhi\ values inferred from the extinction, using the average $\ebv/\nh$ ratio. 
These values are not suitable for our analysis, as we need \textit{independent} dust extinction and gas column density measurements, to probe the \ebv-\nh\ relationship and variations in \nhav.

More accurate \nhi\ measurements were made using spectroscopic data and a continuum reconstruction method for the Ly$\alpha$ absorption line \citep{1975ApJ...200..402B,1994ApJ...427..274D,2003ApJ...594..279G}.
If available, Hubble Space Telescope (HST) data with the Space Telescope Imaging Spectrograph (STIS) and the E140H or E140M echelle grating were used, as these were found to provide the best signal-to-noise ratio and resolution compared to the other options.
These data can be found in MAST: \dataset[https://doi.org/10.17909/5qq7-4c81]{https://doi.org/10.17909/5qq7-4c81}.
For most of the stars, the only available UV spectroscopy data were from the International Ultraviolet Explorer (IUE).
For our analysis, we preferred high-resolution IUE data whenever they were available, as these spectra allow us to carefully exclude stellar features.
In some cases, we used low-resolution IUE data, either because no other data were available, or because the high-resolution data were too noisy.
If multiple observations or exposures of the same type were available, the individual spectra were resampled onto a common wavelength grid using nearest neighbor interpolation.
These resampled spectra were then co-added by taking a weighted average at every point of the common wavelength grid.
The weights used at wavelength $\lambda_i$, are $n_i / f_i \times t_\text{exp}$, with $n_i$ the net count rate (counts / s), $f_i$ the flux (flux unit / s), and $t_\text{exp}$ the sum of the exposure times.
In other words, the data are weighed with a factor $\text{sensitivity} \times \text{exposure time}$, because $n_i / f_i$ is the sensitivity (counts / flux unit).
In the case of high resolution spectra, an additional rebinning step was applied, onto a uniformly spaced wavelength grid with a bin width of \SI{0.25}{\angstrom}.
We found that this bin width provides a good balance between noise reduction and resolution, making it easier to visually inspect the spectra and select suitable wavelength windows, and to run the fitting.

The continuum reconstruction method starts with a linear model fit to estimate the continuum level. 
The data points used for this fit were chosen by visually inspecting plots of the spectra, and manually selecting suitable wavelength ranges far enough away from the \lya\ line.
The model for the \lya\ line is given by $F_{\text{model}}(\lambda_i) = F_{\text{continuum}}(\lambda_i) \phi(\lambda_i)$, where the first factor is the linear continuum model, and $\phi(\lambda)$ is the absorption profile of the \lya\ line as given by \citet{1994ApJ...427..274D}.
This model has one parameter, $\log \nhi$, which is optimized utilizing a least squares approach to minimize the difference between the observed and model flux for a set of carefully selected data points. 
The data points chosen for this are preferably in the \lya\ wings, and not contaminated by any large stellar or interstellar spectral features.
The approach for choosing wavelength points, fitting the continuum, and fitting the \lya\ line, is analogous to the one presented in \cite{2019ApJ...871..151R}, barring a correction for the line-of-sight velocity.

\begin{figure*}
    \centering
    \includegraphics{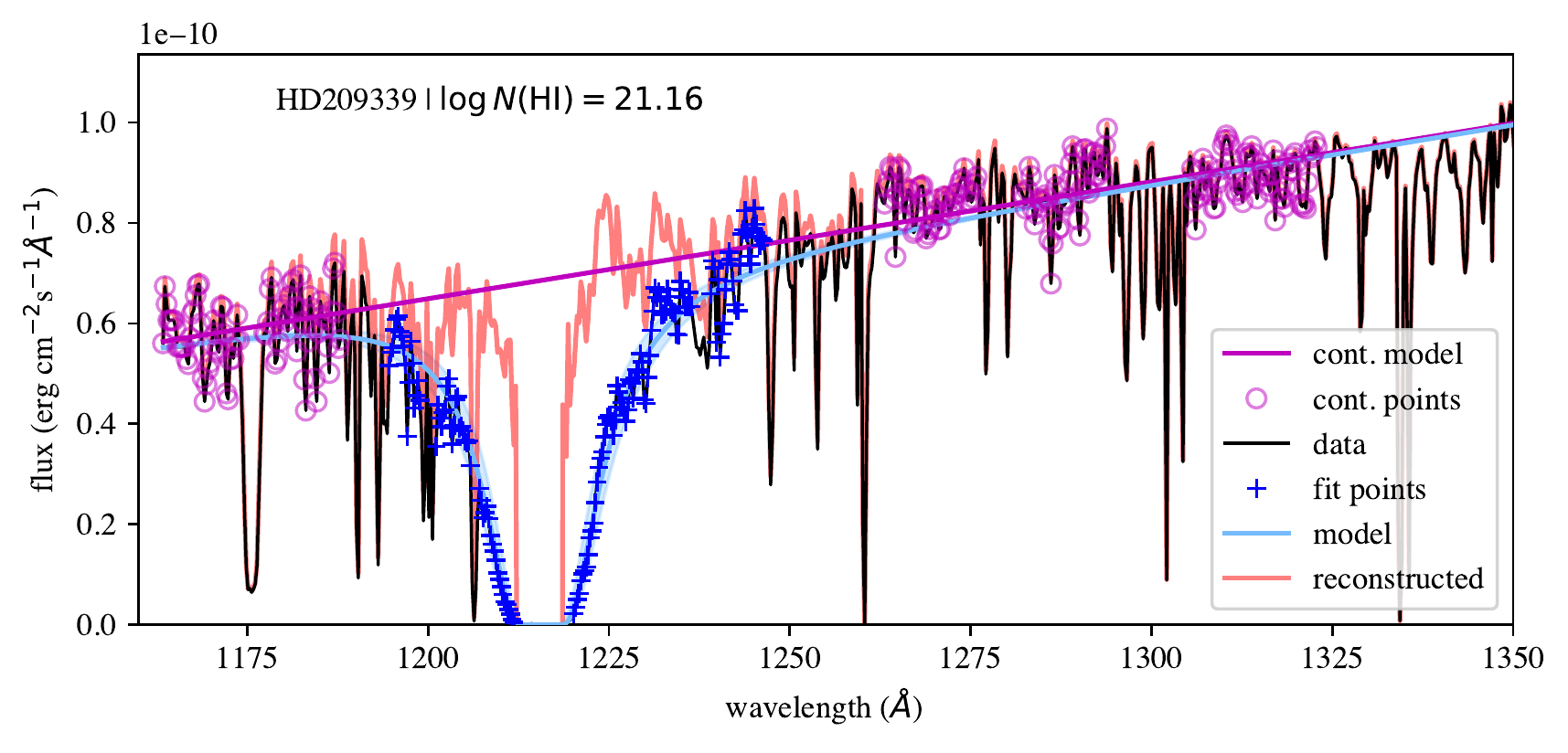}
    \caption{Example \lya\ absorption profile fit. 
    \textit{Black:} ancillary data from HST/STIS.
    \textit{Magenta:} data points selected to fit the continuum, and the resulting linear continuum model. 
    \textit{Blue:} data points selected for \lya\ wings, and the resulting line profile.
    \textit{Red:} Data with \lya\ profile divided out, which brings those data back to the continuum level. This is used as a visual aid to judge the quality of the fit.}
    \label{fig:lya}
\end{figure*}

The noise on the model flux $\sigma_c$, is an estimate for the noise due to both observational and instrumental effects, and smaller spectral features which were not excluded from the fitting range.
It is assumed constant over the entire wavelength range, and is estimated by taking the standard deviation of $F_{\text{data}}(\lambda_i) - F_{\text{continuum}}(\lambda_i)$, where $i$ runs over the same wavelength range that was used for the continuum fit.
A demonstration of the wavelength range selection and the resulting absorption line model is shown in Figure~\ref{fig:lya}.

The uncertainties shown for $\log \nhi$ in Table~\ref{tab:gasdetails}, are half of the distance between the 16th and 84th percentiles of the likelihood function $L(\log \nhi ) \propto \exp \chi^2$.
This value for the uncertainty depends on the noise estimate $\sigma_c$, but should be a reasonable estimate for most data points.

\subsection{Sightline Properties}
\label{sec:properties}

We extended the GCC09 data with accurate measurements of the atomic and molecular column densities \nhi\ and \nhtwo, so that the total column \nh, the gas-to-dust ratio \nhav\, and the molecular fraction \fhtwo\ can be derived.
The extinction curves themselves, and the procedures used to construct them from the raw data and hydrogen abundances, may be found in GCC09.
The GCC09 data contains a set of parameters describing the shape of each extinction curve, together with the optical extinction quantities \av\ and \rv\ as derived from broadband photometry.
The extinction curve shape is represented by the six FM90 \citep{1990ApJS...72..163F} parameters, which are found by fitting the measured extinction for each wavelength $E(\lambda-V)/E(B-V)$ with the functional shape given by FM90.
Note that we use slightly different parameters, because GCC09 fitted the extinction curve model to $A(\lambda) / A(V)$, instead of the customary $E(\lambda-V)/E(B-V)$.
Aside from this minor distinction, the model equation remains identical to Equation 2 of \citet{1990ApJS...72..163F}.

\begin{figure}
    \centering
    \includegraphics{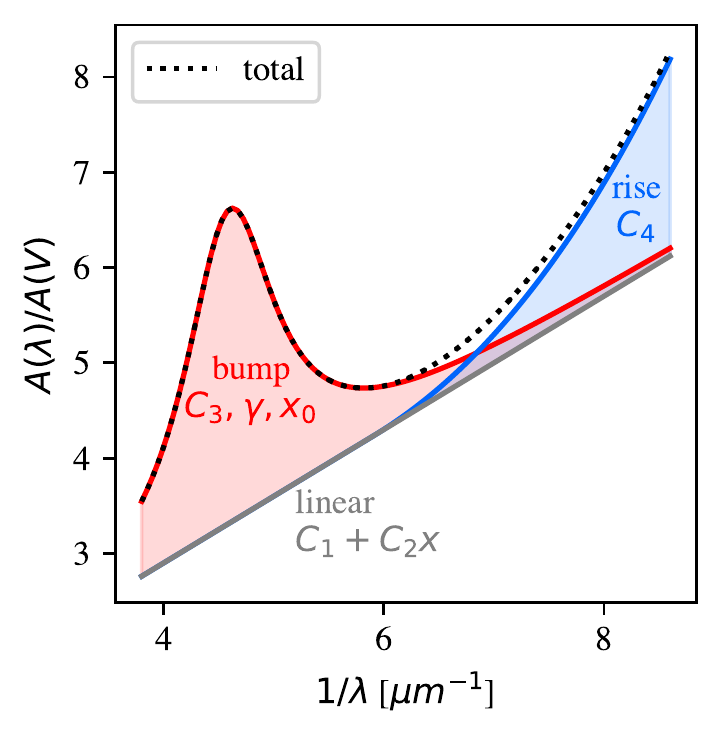}
    \caption{Sketch of the FM90 extinction curve model \citep[Equation 2]{1990ApJS...72..163F}, with arbitrary parameters.
    The functions describing the \SI{2175}{\angstrom} bump and the FUV rise are added on top the linear part of the extinction model.}
    \label{fig:extinction}
\end{figure}

We denote the FM90 parameters as \cav{1}, \cav{2}, \cav{3}, \cav{4}, $x_0$, and $\gamma$.
Each parameter corresponds to a certain aspect of the extinction curve.
As illustrated in Figure~\ref{fig:extinction}, the parameters $\cav{1}$ and $\cav{2}$ describe the linear portion of the curve.
The nonlinear features added to this linear base are the FUV rise ($\cav{4}$), and the \SI{2175}{\angstrom} bump ($\cav{3}$), for which $x_0$ is the central wavelength and $\gamma$ is the width.
The bump amplitude can be derived from these parameters, and is equal to $\cav{3}/\gamma^2$ \citep{1990ApJS...72..163F}, and the area of the bump is $\pi \cav{3}/2 \gamma$.
By evaluating the functional form of the extinction curve using the fitted FM90 parameters, we can obtain extinction ratios at specific UV wavelengths.
We use this to focus on the UV-to-optical extinction ratio \akav\ in the range of the the FUV rise, and \abumpav\ at the bump center (wavelength between parenthesis in \AA), in addition to the optical quantity $\rv$.

Next to the \nhi\ and \nhtwo\ data measured and collected in this work, Table~\ref{tab:gasdetails} also lists a few derived quantities, such as the total gas column density and the molecular fraction.
They are defined as $\nh = \nhi + 2\nhtwo$, and $\fhtwo = 2\nhtwo / \nh$.
For completeness, we also calculated a set of rough estimates for the total number density $n(\text{H})$ (\si{\per\cubic\cm}), by dividing the calculated column densities by the distance $d$.
Photometric distance values from \citet{2021ApJ...911...55S} were used for the 39 stars common with our sample.
For the other stars, the distances were derived from parallax values provided by Gaia DR2 \citep{2018A&A...616A...1G}.

The rotational temperature $T_{01}$ of H$_2$ is derived using the Boltzmann equation and the ratio of the column densities of H$_2$ in its first two rotational states $J = 0$ and $J = 1$ \citep{2009ApJS..180..125R}.
We use $N(0)$ and $N(1)$ from Table \ref{tab:h2dat}.
\begin{equation}
  T_{01}=\frac{74.0}{\log N(0)-\log N(1)+0.954}.
\end{equation}
When the balance between the populations of these states is driven by thermal collisional transitions, then the rotational temperature is a good measure for the kinetic temperature of the gas where H$_2$ is present.

\subsection{Uncertainties and Covariance}

\subsubsection{Covariance Estimation for Derived Quantities}

In certain plots of the results presented in Section \ref{sec:results}, relationships between two derived quantities are shown, some of which depend on a common factor.
A common factor can introduce correlations between the quantities shown on the x and y axes.
If the uncertainty in the common factor is large, the resulting correlations might affect our conclusions about the sample.
To be clear, this section concerns correlations between different quantities for the same sightline due to the nature of the measurements, and not correlations between different sightlines.

To take these correlations into account, an estimate for the covariance is needed.
We calculate the covariance between two derived quantities using the standard, first order expansion method.
For three stochastic variables $x, y, z$, and two functions $f(x, z)$ and $g(y, z)$, which both have a dependency on $z$, the covariance for a data point $x_i, y_i, z_i$ can be estimated using
\begin{equation}
  \label{eq:covariance}
  \text{cov}(f(x_i, z_i), g(y_i, z_i)) = \sigma^2_{z,i} \frac{\partial f}{\partial z}(x_i, z_i) \frac{\partial g}{\partial z}(y_i, z_i).
\end{equation}
As a practical example, we explain the situation when \nhav\ and \rvi\ are compared.
In the work by GCC09, \av\ was derived by extrapolating photometric measurements of $E(J-V)$, $E(H-V)$, and $E(K-V)$.
We therefore consider \av\ and \ebv\ to be independent, while the \rv\ measurement is not independent because it was derived using $\rvi = \ebv / \av$.
The common factor between \nhav\ and \rvi\ is \av, so the covariance is given by
\begin{equation}
  \label{eq:cov.nhav.rvi}
  \text{cov}\left(\frac{\nh}{\av}, \rv^{-1}\right) = \nh \ebv \frac{\sigma_{\av}^2}{\av^4}.
\end{equation}
In this example, \nh\ is independently measured from \av\ and \ebv, so it does not contribute a correlation term. 
Technically, \av\ and \ebv\ are correlated because $\ebv = A(B) - \av$, but since the uncertainty is dominated by $A(B)$, we consider them to be uncorrelated for this example.

In certain plots of this work, the standard deviations and covariances calculated via this method are visualized using ellipses.
The length and orientation of the axes of each ellipse was determined by calculating the eigenvalues and eigenvectors of the covariance matrix for each point, given by
\begin{equation}
  \label{eq:matrix}
  \mathcal{C}_i =
  \begin{pmatrix}
    \sigma^2_{x,i} & \text{cov}(x_i, y_i)\\
    \text{cov}(x_i, y_i) & \sigma^2_{y,i}
  \end{pmatrix}.
\end{equation}
where the elements were obtained using the standard error propagation methods described above.

\subsubsection{Linear Fitting with Covariance}
\label{sec:slopefitting}

We report the best fitting slope and intercept of a linear model for certain parameter pairs.
In all cases, the measurement uncertainties are substantial for both x and y, and in one case, they are correlated.
A simple weighted regression with error bars in the y-direction cannot be used, as it would ignore the uncertainties in the x-direction, and the xy-correlations.
To include these aspects of the uncertainty in the calculation of the best fitting line, a likelihood function was set up which is based on the perpendicular distance of each data point from the linear model.
From the xy-covariance, an uncertainty in the direction perpendicular to the line is derived by performing a projection onto the normal of the line.
This perpendicular uncertainty determines how many standard deviations a data point deviates from the linear model.
The reasoning and equations behind this likelihood function are given in \cite{2010arXiv1008.4686H} and \citet{2015PASA...32...33R}.

The best fitting line is obtained by maximizing this 2D likelihood function, $L(m, b)$, where $m$ and $b$ are the slope and intercept of the line.
To estimate the uncertainty on these parameters, the likelihood function was evaluated on a grid of points around the solution, and then used as a probability density function to draw 2000 random $(m, b)$ pairs.
The uncertainties reported in this work are the standard deviations of these sampled $m$ and $b$ values.

\subsubsection{Sample Correlation Coefficient and Significance}
\label{sec:pearson}

For every scatter plot in this work, quantify to what degree the observed quantities are correlated, and how significant the correlation is.
For this reason, we report values for the Pearson correlation coefficient $r$, and a level of significance expressed as a number times a certain $\sigma$.
The paragraphs below clarify how we calculate these indicators while including the general xy-uncertainties described above.

\paragraph{Uncorrelated Uncertainties} The significance level of a certain value for $r$ is defined as the probability that an uncorrelated dataset produces
a value of $r$ which is as large as the one observed, solely because of statistical fluctuations.
In a previous study of the correlations between \fhtwo\ and the FM90 parameters by \cite{2002ApJ...577..221R} the significance was expressed as a numerical multiplier times a certain $\sigma$, e.g.\ ``$3.4\sigma$''.
When determining the significance that a correlation coefficient is nonzero, this $\sigma$ relates to the $r$-distribution under the null-hypothesis that the underlying physical data are fully uncorrelated.

The commonly used analytical equations for the significance of $r$ only work when the underlying distribution of the data is bivariate normal.
Because the sightlines in our sample were selected only if a high-quality extinction curve could be obtained, the distribution for the measured quantities is non-normal.
Therefore, we perform a random permutation test instead, a resampling technique commonly used for dealing with non-normal data.
Permutation testing is a general technique for testing hypotheses that relate to certain orderings or associations of the data points. 
The main principle is choosing a test statistic (such as $r$), and seeing how it behaves when the labels of the data are scrambled.
See the book by \citet{good1994} or blog posts (e.g.  \url{https://towardsdatascience.com/bootstrapping-vs-permutation-testing-a30237795970}) for use cases and examples.
In this case, a non-zero value for $r$ is expected if the measurements $x_i$ and $y_i$ each belong to the same sightline, and the underlying physics cause the quantities $x$ and $y$ to be correlated.
Scrambling the labels turns the set of dependent measurements $x_i, y_i$, into one of independent measurements $x_j, y_i$, thereby breaking the correlation, resulting in a population that has $\rho = 0$ by construction, or an $r$ distribution centered around 0 when this population is sampled.

For each pair of parameters displayed in one of the scatter plots, we take the effect of the uncertainties on $r$ into account by redrawing the scrambled data sets many times in a Monte Carlo fashion.
Each of the 75 data points is redrawn from a bivariate normal distribution, of which the parameters are defined according to the estimated covariance matrix for that point.
A sample of 6000 $r$-values is created by calculating the regular Pearson correlation coefficient for each realization of the dataset generated this way.
This results in a sample of coefficients $r_0$, centered around zero, and the sample standard deviation $s(r_0) = \sqrt{\sum_i r_{0,i}^2}$ is the value we use to express the confidence as a number of $\sigma$.
The displayed significance level in the corner of each plot is then $r_\text{measured} / s(r_0)$.

\begin{figure}
    \centering
    \plotone{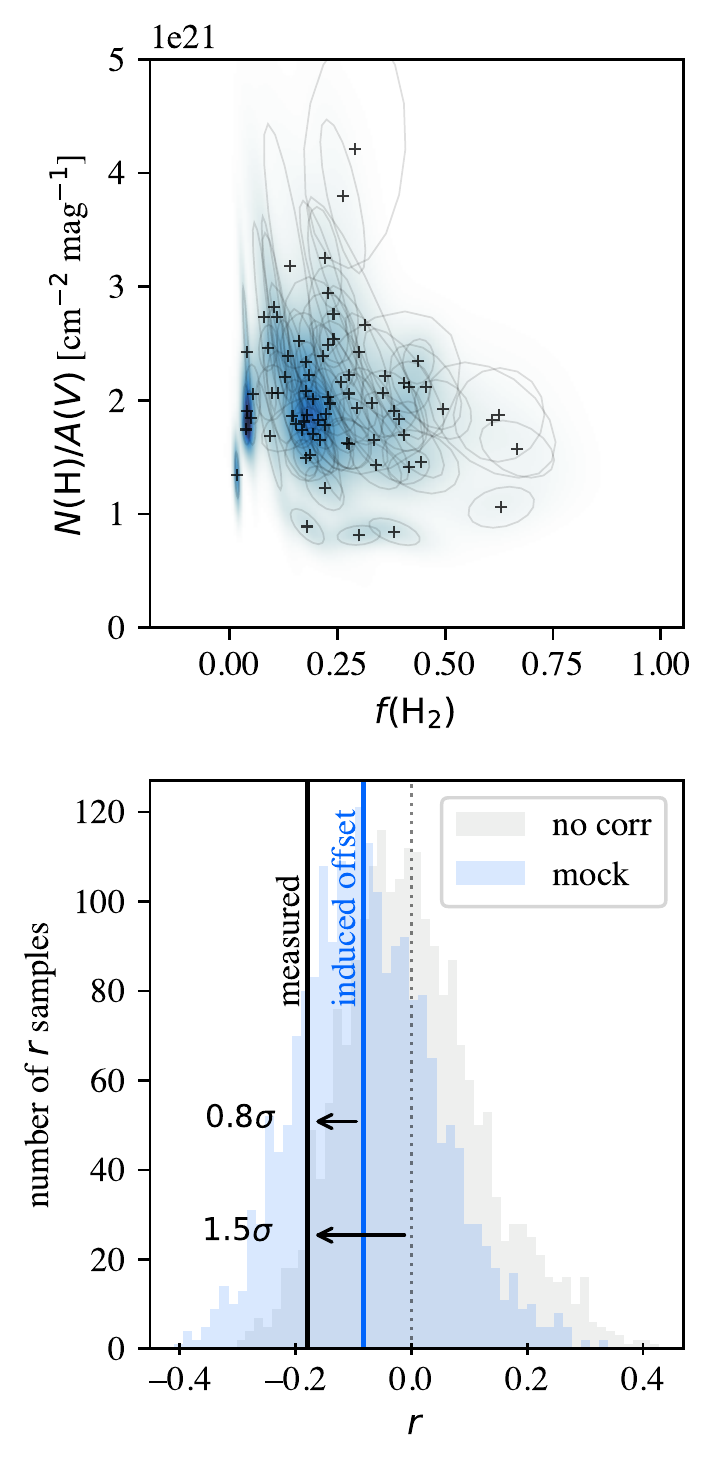}
    \caption{Demonstration of our method for dealing with correlated uncertainties. 
    The top panel shows a scatter plot between \nhav\ and \fhtwo, and the ellipses and density map visualize the correlations due to the common \nhi\ and \nhtwo\ factors in both quantities.
    The gray and blue histograms show the sample of correlation coefficients $r$ before and after applying  mocked correlated measurement errors to the scrambled samples.
    The two arrows and the displayed $\sigma$ values, show how the significance level of the measured correlation changes, when the offset induced by the correlated measurement uncertainties is taken into account.}
    \label{fig:hist}
\end{figure}

\paragraph{Correlated Uncertainties}
Even when the underlying physical data are uncorrelated, a strong correlation in the measurement errors of x and y will lead to a $r_0$ distribution that is no longer centered around zero. 
We call this bias an \textit{induced} correlation, as it is intrinsic to the correlated measurements, and not correlations in the underlying physics.
To adjust the significance level, we need to generate a biased sample of $r_0$ values, so that the expression for the significance becomes $[r_\text{measured} - \text{median}(r_0)] / s(r_0)$.

To generate this sample, it is crucial that the simulated (correlated) measurement noise is applied \textit{after} the scrambling step described in the method above.
This generates a mock observation of the sample, where the measurement noise induces a non-zero correlation into the inherently uncorrelated (scrambled) sample.
Generating an appropriate shift for each data point, to simulate the measurement noise, is not trivial, because the scrambling makes it unclear what should happen with the covariance matrix for each data point.
As an approximation, we randomly assign the covariance matrices to the data points, and then draw the simulated measurement noise according to the corresponding bivariate normal distributions.
The shift in the average of $r_0$, induced by adding this mock data for the correlated measurement noise, is demonstrated in Figure~\ref{fig:hist}.

\section{Results}
\label{sec:results}

\subsection{Gas Column vs.\ Dust Extinction}
\label{sec:column}

\begin{figure*}
  \plotone{{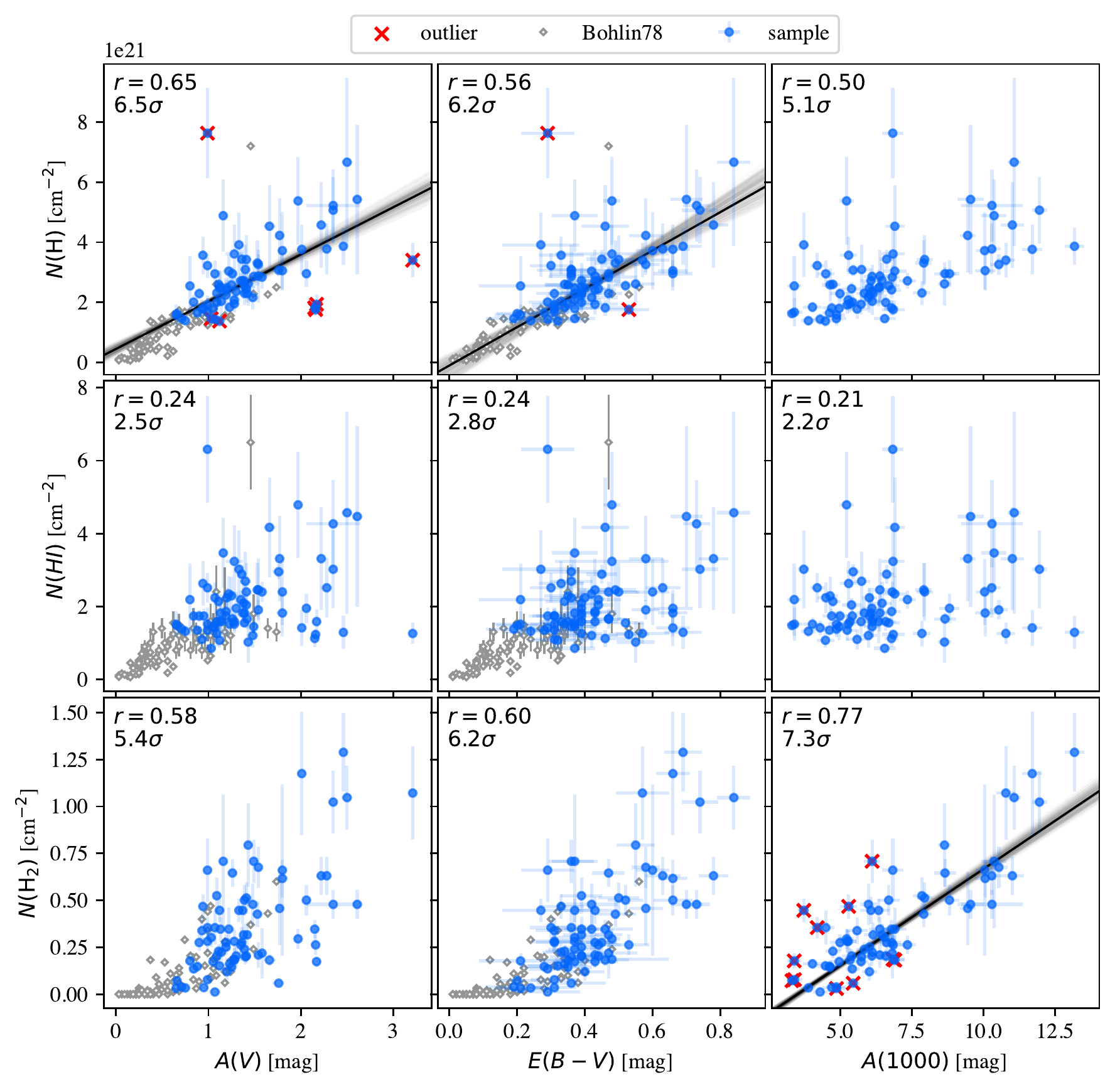}}
  \caption{Total hydrogen, atomic hydrogen, and molecular hydrogen column densities vs.\ dust column, measured via the extinction \av\, \ebv, and \ak.
    Our sample of 75 stars is shown using blue circles, and the error bars represent the $1\sigma$ uncertainty on the measurements.
    The original \textit{Copernicus} sightlines by \citet{1978ApJ...224..132B} are shown for reference, using the grey diamonds.
    The black lines visualize the fit results from the method described in Section \ref{sec:slopefitting}, with the gray auras showing the uncertainty on each fit.
    The numbers in each corner show the Pearson correlation coefficient of the data and its significance, as explained in Section \ref{sec:pearson}.
    Data that deviate more than $3\sigma$ from the fitted line, are indicated with a red `x'.
    They have been excluded from the fit for \nh\ vs.\ \av, but not from the fit for \nhtwo\ vs.\ \ak. 
    In the latter plot, they are only highlighted to show the intrinsic width of the \nhtwo\ vs.\ \ak\ relationship.}
  \label{fig:columns}
\end{figure*}

\subsubsection{Total Column -- Optical Extinction}
\label{sec:colcol}

The correlations between the gas column densities \nh, \nhi, and \nhtwo, and the dust extinction in the V-band \av\ and in the UV \ak, are shown in Figure~\ref{fig:columns}.
The latter extinction was calculated by evaluating the FM90 extinction curve model $A(\lambda) / \av$ at \SI{1000}{\angstrom}.
This wavelength was chosen because it is in the UV absorption/dissociation band of H$_2$.
Since we are taking this value from a fitted, smooth extinction model, the data for all wavelengths near \SI{1000}{\angstrom} is taken into account.
Choosing any value between 912 and \SI{1150}{\angstrom} will result in the same conclusions.

The total H column \nh\ correlates best with the V-band extinction \av.
{It is interesting to note that \citet{2021ApJ...911...40B} found that $A(\sim 3000)$ correlates better with \nh, but this does not significanly affect the discussion and conclusions in this work.}
A measurement of the average gas-to-extinction ratio for the sample is obtained by fitting a linear model to the \nh-\av\ relationship.
Because the uncertainties on the data are substantial for both the \nh\ and \av\ axis, we use the least squares linear model described in Section \ref{sec:slopefitting}.
The results for the linear fits performed throughout this paper are summarized in Table\ \ref{tab:fit}.

\begin{deluxetable*}{cc|cc}
\tablecaption{Slope $m$ and intercept $b$ of the best fitting linear model, for several parameter pairs $(x, y)$ displayed in {Figures \ref{fig:columns}, \ref{fig:c4contribution}, and \ref{fig:rvtrend}}. \label{tab:fit}
}
\tablehead{\colhead{x-axis} & \colhead{y-axis} & \colhead{slope $m$} & \colhead{intercept $b$}}
\startdata
           &        & [\SI{e20}{\per\square\cm\per\mag}] & [\SI{e20}{\per\square\cm}]        \\
\av        & \nh    & $15.73 \pm 0.98$                   & $4.5 \pm 1.1$                      \\
\ebv       & \nh    & $63.8 \pm 5.8$                     & $-1.0 \pm 2.3$                     \\
\ak        & \nhtwo & $1.027 \pm 0.041$                  & $-3.60 \pm 0.22$                   \\
\ak$_\text{rise}$ & \nhtwo & $2.20  \pm 0.09$ & $-0.95 \pm 0.12$\\
$A(2175)_\text{bump}$ & \nhtwo & $3.33 \pm 0.17 $ & $-1.76 \pm 0.20$\\
\hline
           &        & [\SI{e20}{\per\square\cm\per\mag}] & [\SI{e20}{\per\square\cm\per\mag}] \\
\rvi       & \nhav  & $101.3 \pm 8.6$                    & $-12.6 \pm 2.6$                    \\
\akav      & \nhav  & $3.19 \pm 0.22$                    & $2.54 \pm 0.97$                    \\
\abumpav   & \nhav  & $9.43 \pm 0.67$                    & $-10.7 \pm 2.0$                    \\
\enddata
\end{deluxetable*}

For \nh\ vs.\ \av\ we find a slope of about \SI{1.6 \pm 0.1}{\per\square\cm\per\mag} which is significantly lower than the average \nhav\ ratio obtained from observations of X-ray sources, which is around \SI{2e21}{\per\square\cm\per\mag} \citep{1973A&A....26..257R, 1975ApJ...198...95G, 1995A&A...293..889P, 2009MNRAS.400.2050G, 2017MNRAS.471.3494Z}.
Examining the work by \citet{2017MNRAS.471.3494Z}, the X-ray data mainly cover the \nh\ range from \num{2e21} to \SI{2e22}{\per\square\cm}, or $\av = 1\text{ to } 10$, while our data are in the \nh\ range between \num{2e21} and \SI{6e21}{\per\square\cm}.
According to the above data, the \nhav\ ratio seems to be systematically lower in the more diffuse column density regime covered by this work, with $\av < 3$, compared to the average over the entire \av\ range from 1 to 10.
Applying our fitting method to \nh\ vs.\ \ebv, yields a value which is slightly higher than, but consistent with the classic gas-reddening ratio $\nhi / \ebv = \SI{5.8e21}{\per\square\cm}$ from \citet{1978ApJ...224..132B}, and similar to the values reported by others since then \cite{2002ApJ...577..221R, 2009ApJS..180..125R, 2021ApJ...911...55S}.

\subsubsection{H$_2$ Column -- UV Extinction}
We expect to see a relationship between \nhtwo\ or \fhtwo\, and the extinction measured at wavelengths within the H$_2$ dissociation band.
Therefore, we have calculated \ak\ from the parameterized extinction curves, and show its correlation with the gas contents in the third column of Figure~\ref{fig:columns}.

Unlike \av\ and the total gas density, the relationship between \av\ and \nhtwo\ exhibits a much larger scatter.
However, in the lower right panel of Figure~\ref{fig:columns}, a well defined relationship between \ak\ and \nhtwo\ is discovered, with a strong linear correlation coefficient ($\sim 0.8$).
Fitting a slope using the same method yields the result in Table~\ref{tab:fit}, and it can be derived that the fitted line intersects the x-axis at $\ak = 3.4$.
This value can be interpreted as a minimum magnitude of UV extinction, required to observe a significant amount of H$_2$ along the sightline, although the marked outliers in Figure~\ref{fig:columns} make it clear that there is still a significant spread on the H$_2$ column density.
For sightlines with $\ak \leq 7$, many data points deviate more than $3 \sigma$ from the best fitting line, showing that there is a significant amount of scatter on this relationship.
For $\ak \geq 7$, the error bars on \nhtwo\ are too large to discern if this scatter persists.

Small grains absorb primarily at short wavelengths, while larger grains also absorb efficiently at longer wavelengths such as in the V-band.
We will refer to the FUV-absorbing population as ``small grains'' in the rest of this work.
As \ak\ does not correlate with \nhi, these small grains appear to be present only in the molecular regime, where they contribute to a large fraction of the FUV extinction.
The relationship between \nhtwo\ and \av\ {has more scatter, because the larger grains also appear in the \hi\ gas.}

\begin{figure}
    \centering
    \includegraphics[]{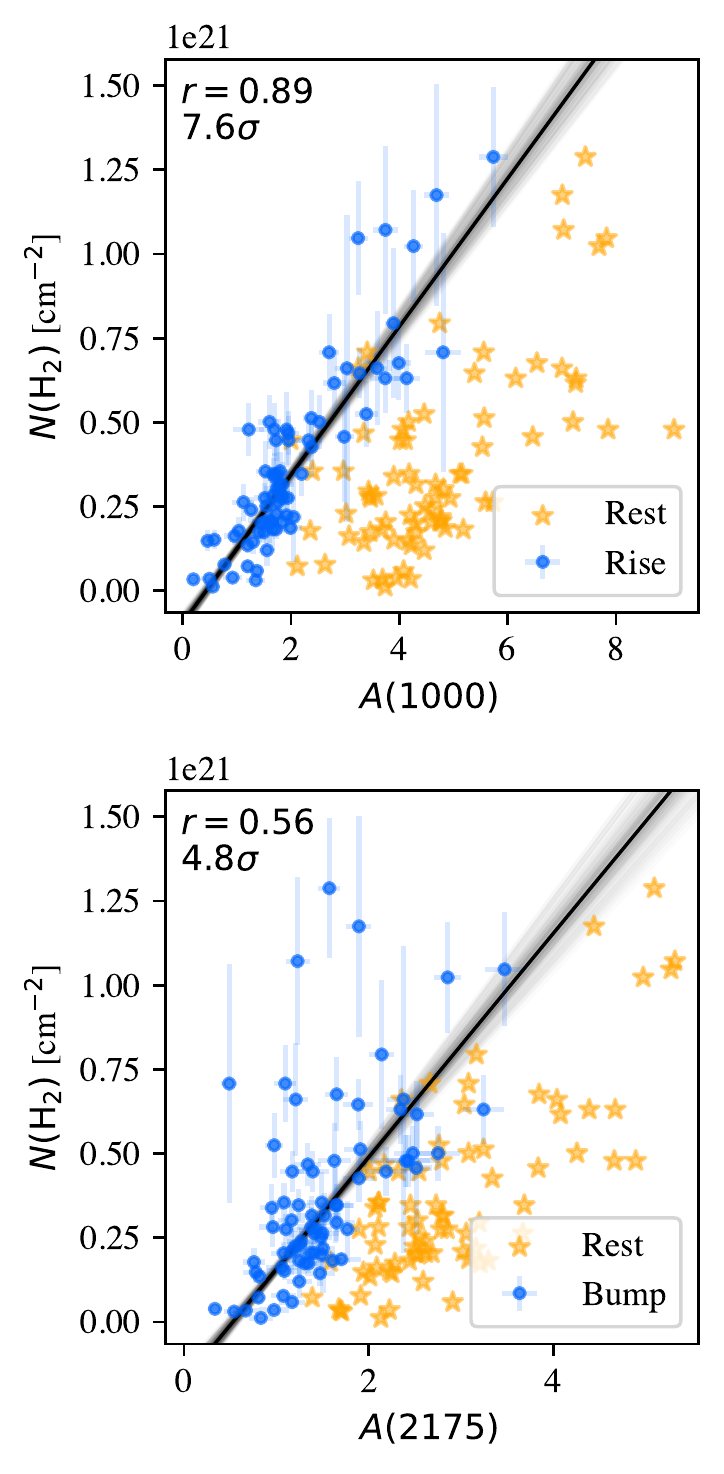}
    \caption{{Upper panel: A different view of the \nhtwo-\ak\ correlation, where we separately show the contribution of the FUV rise $A_\text{rise} = \cav{4} F(x) \av$ (blue circles) and remainder of the extinction $A_\text{rest} = \ak - A_\text{rise}$ (orange stars).
    The observed \nhtwo\ correlation of 0.89 is even higher than the one observed for the total value of \ak.
    The remaining \ak\ extinction still correlates with \nhtwo\, but with a coefficient of 0.63.
    The FUV rise contribution approaches zero when \nhtwo\ is small while the rest does not. 
    Lower panel: Analogous view of the extinction at \SI{2175}{\angstrom}, separating the contribution of the bump from the rest.
    While the bump contribution still nears zero at small \nhtwo, it has a weaker correlation of only 0.56. The rest contribution correlates the strongest in this case, with $r = 0.75$.
    The fit results of both panels are shown in Table \ref{tab:fit} }}
    \label{fig:c4contribution}
\end{figure}

{The FM90 extinction curve model, gives us a way to separate the contribution to \ak\ by the FUV rise specifically, from the total extinction.
The \textit{absolute} contribution by the FUV rise is given by $\cav{4} F(x) \av$, with $F(x)$ as defined in \citep{1990ApJS...72..163F}.
By using this quantity in Figure \ref{fig:c4contribution}, we find a correlation with \nhtwo\ which is even stronger, at almost 0.9.
On the other hand, the FUV rise term contributes to at most 40\% of the total FUV extinction.
After subtracting $A_\text{rise}$, the residual extinction shown as the `rest' in Figure \ref{fig:c4contribution}, still has a noticeable correlation of 0.63, but it is also the main source of the scatter on the \ak-\nhtwo\ relationship.
Most notably, the FUV rise contribution is consistent with a relationship that intersects the origin, while the residual contribution is not. 
For both components, no correlation with \nhi\ was observed.
These findings impose a strong constraint on the nature of the FUV rise carrier: In the diffuse Galactic ISM, it only resides in molecular gas.}

{An analogous approach was taken to investigate the absolute contribution of the NUV bump to $A(2175)$.
For this component, the results are not as clear-cut.
In fact, the `rest' term exhibits a stronger correlation with \nhtwo\ than the `bump' term $A_\text{bump} = \cav{3} \gamma^{-2} \av$: the coefficients are 0.75 and 0.56 respectively.} 
{
Given this difference in behavior between the FUV rise and NUV bump, it is unlikely that both of these features are produced by the same carrier.
However, it remains possible that the two carriers have a common origin in terms of physical conditions and formation processes, considering that the behavior with respect to \nhtwo\ is somewhat similar at low column density.
At higher column densities, it seems that the NUV bump amplitude stops increasing, while that of the FUV rise keeps following the trend.}
Further interpretation and possible causes for the observed relationships are discussed in Section\ \ref{sec:smallgrainh2}.

\subsection{Gas-to-dust Ratio vs.\ Dust Extinction Ratios}

\begin{figure*}
  \centering
  \plotone{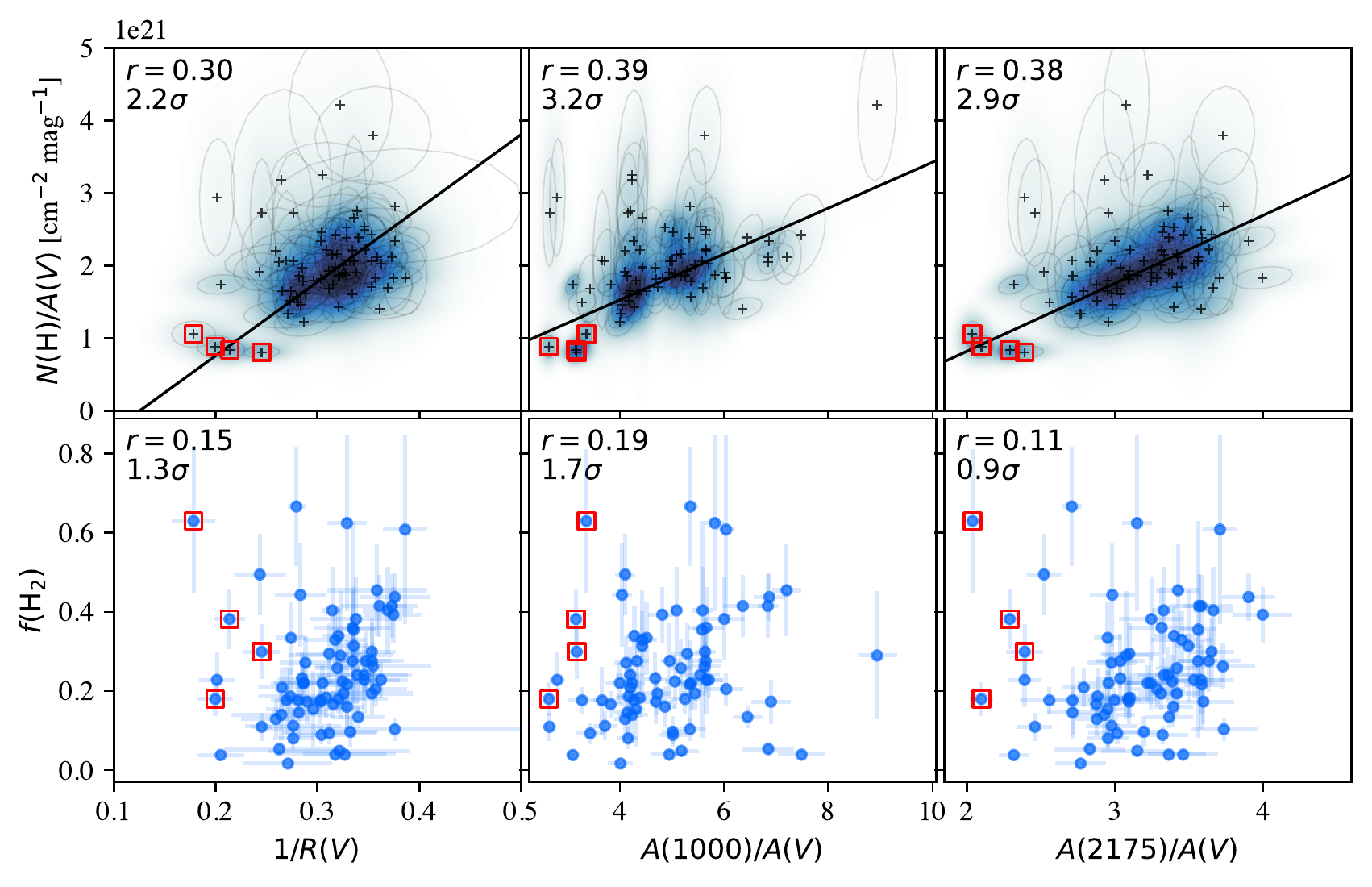}
  \caption{Gas to dust ratio and molecular fraction versus three extinction ratios.
    The four low-\nhav\ sightlines towards HD045314, HD164906, HD200775, and HD206773 have been indicated with red squares.
    HD096675 has an exceptionally high value for \nhav, and has been excluded from these plots and the analysis.
    Due to the common factor \av\ for the quantities shown on the axes in the top row, the uncertainties are correlated in x and y.
    The shapes of the correlated uncertainties are visualized by the ellipses, and the color map shows where the data are concentrated, by showing each point as a 2D Gaussian probability density.
    The correlations in the uncertainty are generally small, and do not strongly affect the conclusions.
    The scrambling and resampling test described in Section \ref{sec:pearson} estimates that the correlation coefficient is boosted by about 0.03 by the correlated x and y measurement uncertainties.}
  \label{fig:rvtrend}
\end{figure*}

The \nhav\ data have a weighted average of \SI{1.6e21}{\per\square\cm\per\mag}, which is equal to the fitted \nh-\av\ slope of Section \ref{sec:colcol}.
There are variations in this ratio, but for most of the quantities we measured, we find no clear trend with the variations in \nhav.
Most notably, \nhav\ does not seem to show any systematic variation with the gas properties  \fhtwo\ or $T_{01}$.
{The} top panel of Figure~\ref{fig:hist}, which demonstrated our method to determine the significance of the observed correlations, shows this result for \nhav\ vs.\ \fhtwo.
For the rest of these negative results, we do not show plots in this paper.

There is no systematic variation of \nhav\ with the column density \nh\ or number density \denh.
This result was to be expected, as earlier studies of the dust-to-gas ratio showed that most of the variations occur below $\log \nh = 21.5~[\si{\per\square\cm}]$ \citep{2021ApJ...910...95R}.
Above this density, most of the metals have been converted into dust mass, and the dust-to-gas ratio as a function of column density forms a plateau (Clark et al.\ \textit{in prep}).
There are also no significant correlations of \nhav\ with the dust columns \av\ or \ak, or the FM90 bump parameters $\gamma$ and \cav{3}, or the FUV rise parameter \cav{4}.
However, we did discover a relationship between the gas-to-dust ratio and certain quantities related to the extinction curve shape.

The top row of Figure~\ref{fig:rvtrend} compares \nhav\ with three relative measures of the extinction: $\rvi = \ebv / \av$, \akav, and \abumpav.
There is a positive correlation between \nhav\, and both \akav\ and \abumpav\ at around the $3 \sigma$ level, meaning that sightlines with a stronger UV extinction compared to the optical, also have a lower amount of optical extinction per gas unit.
For the optical quantity \rvi, there seems to be a similar trend when judged with the human eye, but the error bars are too large to have a significant correlation according to our metric.
In Figure~\ref{fig:rvtrend}, the four points corresponding to HD045314, HD164906, HD200775, and HD206773 have been indicated with red squares, because they have a markedly low \nhav\ ratio.
The best fitting slope for \nhav\ was calculated relative to the three parameters \rvi, \akav, and \abumpav, and each time the resulting line goes right through this group of four points.
Even when these low-\nhav\ points are excluded from the data used for fitting, the fitted line still passes through that group.
In other words, their \rvi\ values deviate from the Galactic average of $1 / 3.1 = 0.32$, in a manner that is consistent with a trend fitted to the rest of the sample.
The ratios \akav\ and \abumpav\ show similar behavior, which is expected due to their connection with \rvi, shown in Figure 7 of \citet{2009ApJ...705.1320G}.
The observed trend could be interpreted as the effect coagulation, which increases the amount of dust contributing to \av\ (decreasing \nhav), while reducing the relative amount of small grains (decreasing \akav).

The average \nhav\ of our sample is lower than the Galactic average (see Section \ref{sec:colcol}), because of the sample selection, and the corresponding \nhav\ distribution. 
Instead of being statistically representative of the entire population of Milky Way sightlines, the sample focuses on including a broad range of \rv\ values.
By evaluating the fit for \nhav\ vs.\ \rvi\ at the Galactic average $\rvi = 0.32$, we can adjust the average \nhav\ of our sample for the difference in \rv, and obtain an estimate for the average of \nhav\ as if the average \rv\ were equal to 3.1.
With the values shown in Table\ \ref{tab:fit}, the result is $\langle \nhav \rangle = 0.32m + b = \SI{1.99 \pm 0.05 e21}{\per\square\cm\per\mag}$, and the error on this result was calculated by taking the standard deviation of $0.32m + b$, over a sample of $(m, b)$ pairs that was drawn as explained in Section \ref{sec:slopefitting}.
This value is consistent with the literature value of $\sim \SI{2e21}{\per\square\cm\per\mag}$ \citep{1978ApJ...224..132B, 1994ApJ...427..274D, 2017MNRAS.471.3494Z}, showing that the difference in \rv\ between our sample and the total Galactic sightline population also explains the difference in \nhav.

\subsection{H$_2$ Fraction vs.\ Extinction Curve Shape}

\begin{figure*}
  \centering
  \plotone{{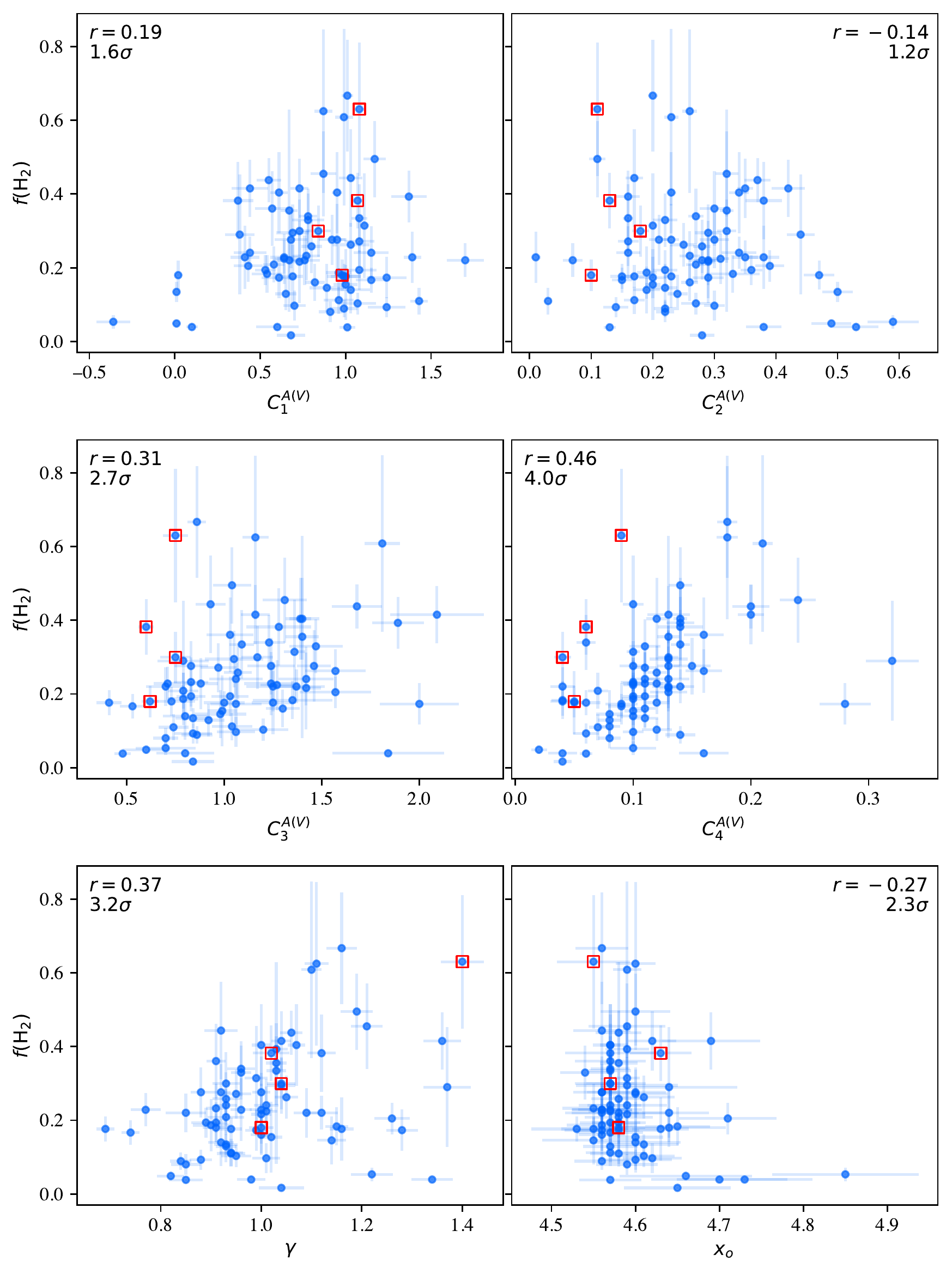}}
  \caption{FM90 parameters of the extinction curves versus molecular fraction \fhtwo.
  The numbers shown are analogous to those in Figure~\ref{fig:columns}
    The four low-\nhav\ sightlines towards HD045314, HD164906, HD200775, and HD206773 have been indicated with red squares, showing that the NUV and FUV feature strengths \cav{3}\ and \cav{4}\ are weak.}
  \label{fig:fh2.vs.params}
\end{figure*}

The sightlines have a wide variety of molecular fractions, ranging from 0.02 to nearly 0.7, with most above 0.1.
The bottom row of Figure~\ref{fig:rvtrend} shows how the molecular fraction \fhtwo, relates to \rvi\ and two UV-to-optical extinction ratios.
The plot for \abumpav\ has a structure similar to the one with \rvi\, while the cloud of data points look somewhat different for \ak.
This is expected as GCC09 already showed that \abumpav\ strongly correlates with \rvi, while the \rv-dependent relationship for \ak\ shows more scatter.
Earlier works which examined \fhtwo\ vs.\ \rv\ found a possible correlation, but it was of limited significance \citep{1988ApJ...335..177C, 2002ApJ...577..221R, 2009ApJS..180..125R}.
{For example, in \citet{1988ApJ...335..177C}, \nhtwo\ was measured for a number of diffuse ($\av < 2$) sightlines, and \fhtwo\ was found to decrease going from $\rv \sim 3$ to $\rv \sim 4$.}
For our data, the calculated correlation coefficient for \fhtwo\ is very weak for all three extinction ratios, because there are many points that deviate from the main cluster of points centered around $\rvi = 0.3$.
It should also be noted that the four low-\nhav\ outliers, again indicated with red squares, lie distinctly separate from the main group of points, with a wide range of \fhtwo\ values.
{By excluding these points from the data, the correlation coefficient increases to 0.3, with a significance of 2.9 for \fhtwo\ vs.\ \rvi\ and 2.3 for both UV-to-optical extinction ratios.
Whether this correlation is observed, therefore depends on the subset of the ISM being probed.}

In addition to the total extinction in Figure~\ref{fig:columns}, and the extinction ratios in Figure~\ref{fig:rvtrend}, we also compare the specific features from the FM90 parameterization with the molecular contents.
In Figure~\ref{fig:fh2.vs.params}, we present plots comparing \fhtwo\ to the six FM90 parameters: \cav{1}, \cav{2}, \cav{3}, \cav{4}, $x_0$, and $\gamma$.
The parameters describing the linear aspect of the extinction curve, $\cav{1}$ and $\cav{2}$, do not show any trends with \fhtwo, nor does the center of the FUV bump $x_0$.
The main trends in \fhtwo\ relate to the parameters for the magnitude and width of the \SI{2175}{\angstrom} bump (\cav{3}, $\gamma$) and the strength of the FUV rise (\cav{4}).
{Considering the results for the column density presented in Section \ref{sec:column}, these results are just a different view of the same relationship.
However, this view allows us to compare these results to previous work.}
Similar results were found by \citet{2002ApJ...577..221R}, who found a correlation at the $2.6 \sigma$ level for $c_4$, and one of $3.7 \sigma$ with $\gamma$.
In this work, the correlation with \cav{4} is the more significant one at $3.3 \sigma$, while the one with $\gamma$ is at $2.7 \sigma$.

As in the other figures, the red squares in Figure~\ref{fig:fh2.vs.params} indicate the same four low-\nhav\ sightlines, which were found grouped together in the \nhav\ vs.\ \rvi\ plot of Figure~\ref{fig:rvtrend}.
In this case, it is interesting that their \cav{3} and \cav{4} parameters are on the lower side of the sample, while their molecular fractions are very different.
So while the dust in these four sightlines seems to have evolved in a similar way, the observed molecular fraction does not seem to play a role in this phenomenon.

\subsection{Density and Rotational Temperature}

\begin{figure}
  \centering
  \plotone{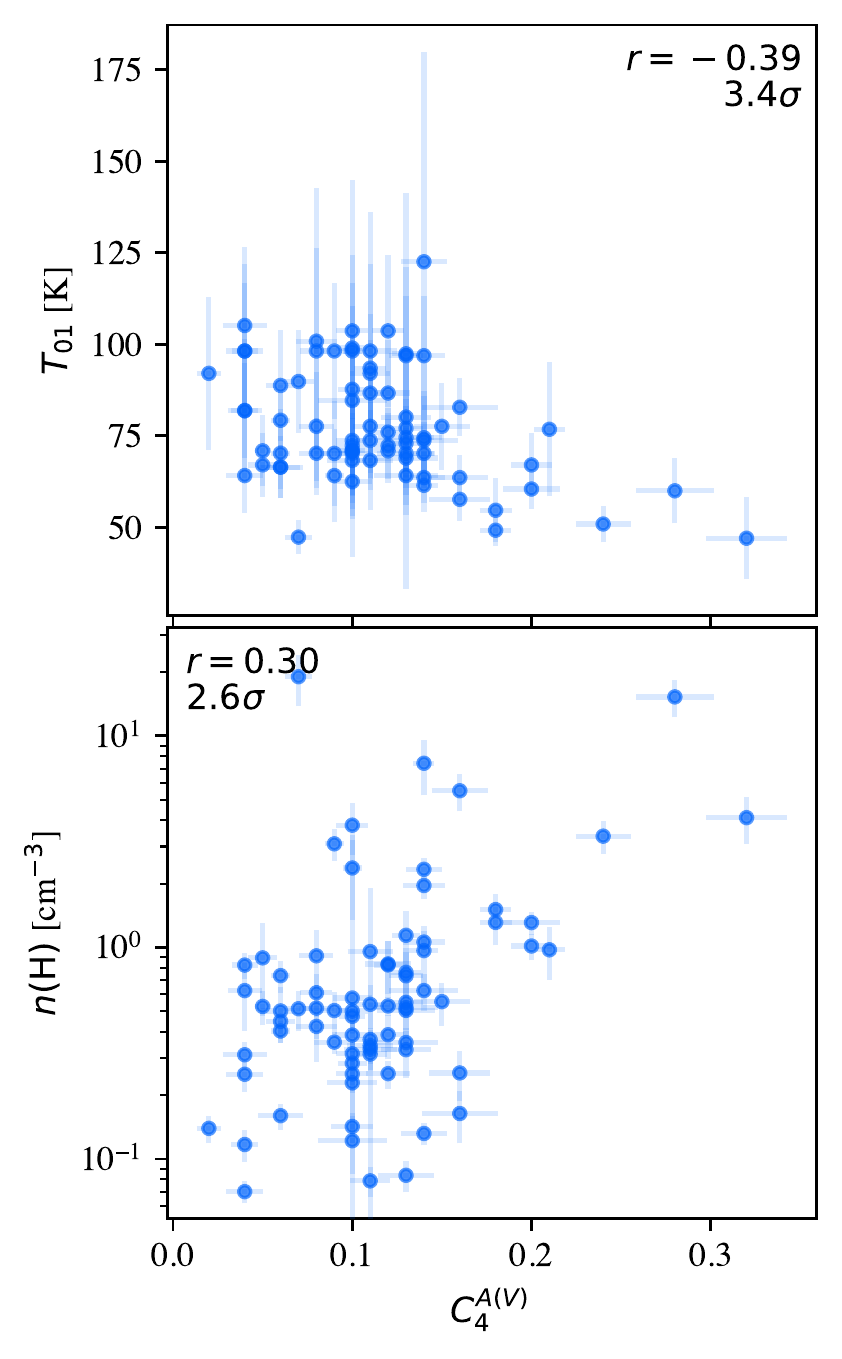}
  \caption{
  Left: Rotational H$_2$ temperature $T_{01}$ and average number density \denh, compared to the FUV rise strength \cav{4}.
  The sightlines that have a strong FUV rise ($\cav{4} \gtrsim 0.2$) have low H$_2$ temperatures and high densities.}
  \label{fig:tempdens}
\end{figure}

We investigated a variety of relationships between the extinction properties, the rotational temperature of H$_2$, and the density estimates $\denh = \nh / d$ as calculated according to section \ref{sec:properties}.
\citet{2009ApJS..180..125R} already showed that at higher temperatures, the medium along MW sightlines generally is of lower density and molecular fraction, but with a lot of scatter.
We confirm these findings in our sample.
Sightlines with low $T_{01}$ values also seem to have a wide range of molecular fractions, with most of the points between 0.1 and 0.7.

In the left column of Figure~\ref{fig:tempdens}, we show how \cav{4} behaves under different gas conditions, using these estimated quantities.
The most striking part is that the points with $\cav{4} > 0.15$ all have H$_2$ temperatures on the low end (40 - 80 \si{\kelvin}), and densities on the high end (1 - \SI{10}{\per\cubic\cm}), relative to the rest of the sample.
Meanwhile, low values of \cav{4} occur at a wide variety of rotational temperatures (60 - 120 \si{\kelvin}) and number densities (0.05 - \SI{10}{\per\cubic\cm}).

\subsection{Depletion}
The data by \citet{2009ApJ...700.1299J} have 31 sightlines in common with this work, of which 29 have a measurement of $F_*$, a quantity which describes the overall depletion of metals in the gas, or how much metals are locked up in dust grains.
We find no obvious correlation of $F_*$ with \nhav, nor with \rvi\ or \akav.
Our number density measurements for these targets confirm the linear relationship between $\log \denh$ and $F_*$ that was already shown in \citet{2009ApJ...700.1299J}.
This means that the exchange of metals between the dust and the gas does not play a major role in influencing the variations in \nhav\ or the extinction curve shape.
Instead, processes such as coagulation and shattering are more likely explanations, where only the size distribution changes, leaving the dust mass and composition mostly unchanged.

\section{Discussion}
\label{sec:discussion}

\subsection{Relationship Between Small Grains and H$_2$}
\label{sec:smallgrainh2}
We found that \nhtwo\ has a strong linear correlation with the absolute extinction \ak, {and specifically with the contribution of the FUV rise term to \ak}. 
This means that wherever there is H$_2$, {the carrier of the FUV rise feature} is also present.
On the other hand there is no obvious relationship between \fhtwo\ and the UV-to-optical extinction ratio, but the FUV rise \cav{4} and \SI{2175}{\angstrom} bump width $\gamma$ do show a significant positive correlation.
The rest of this discussion will refer to the carriers of the UV extinction in question as ``small grains''.
This can be justified by for example the models by \citet{2018A&A...611A...5S}, where the $c_4$ parameter correlates strongly with the mass ratio of small to large grains.

{Given that our findings mostly concern the bump and rise features, the small grains in question likely have a carbonaceous composition.}
This could include very small solid state grains (hydrogenated amorphous carbon) and large molecules such as Polycyclic Aromatic Hydrocarbons (PAHs), both of which are viable carriers for these dust observables \citep{2013A&A...558A..62J}.
As shown by laboratory measurements \citep[e.g.][]{1992ApJ...393L..79J, 2010ApJ...712L..16S} and theoretical calculations \citep{2004A&A...426..105M}, the absorption cross section of PAHs typically exhibits several features in the range of the \SI{2175}{\angstrom} bump. 
However, the extinction curve can become smooth if the size distribution of PAHs is sufficiently broad \citep{2012A&A...540A.110S}.
{In the FUV, these cross sections exhibit a similar behavior as the FUV rise}.

{In the next two parts of the discussion, we speculate about two scenarios which could explain the relationship between small grains and H$_2$.}

\subsubsection{Small Grains Enhance H$_2$ Shielding/Formation}

The first scenario is that the presence of small grains could help the formation and/or survival of H$_2$.
{A certain level of shielding is needed for H$_2$ to survive exposure to dissociating far-UV photons.
PDR models with different (spatially varying) grain size distributions were compared by \citet{2007A&A...467....1G}, and the resulting H$_2$ transitions occurred between $\av \sim 0.1$ and $\av \sim 1$, depending on the number of small grains in the model.
At large \rvi\ or large UV-to-optical extinction ratios, the enhanced UV extinction could lower the minimal \av\ required for H$_2$ to survive.
Most entries in our sample have $\av < 2$ along the entire sightline, so the individual clouds intersected by these sightlines have an \av\ of the order 1 or lower.
Therefore, a sufficient quantity of small, FUV-absorbing grains might be needed to allow the H$_2$ transition to occur in these diffuse clouds.
However, Figure \ref{fig:c4contribution} showed that the underlying cause of the observed relationship is specifically the absolute extinction contribution by the FUV rise, as opposed to the total FUV extinction.
If the shielding by dust was the main driver for the observed H$_2$ relation, one would expect the total dust extinction to be the key quantity determining \nhtwo.
Therefore, while dust shielding in the FUV might be important in the ISM, we do not expect it to be the most important effect driving the observed H$_2$ correlation.}

{Another other option for the scenario where dust enhances the H$_2$ abundance, is that these small dust grains significantly enhance the formation of H$_2$.}
The dominant formation pathway for molecular hydrogen is on the surface of grains, and an increased number of small grains could provide a substantial boost to the available surface area.
{Small grains provide the most surface area because of their much larger number densities.
The available surface area provided by a certain grain size scales with $a^2 n(a)$, where $a$ is the grain size and $n(a)$ is the number density of grains which typically scales with $a^{-3.5}$ \citep{1977ApJ...217..425M}.
The surface area $A$ provided by grains of size $a$ then scales as $A(a) \sim a^{-1.5}$.
Now consider a grain evolution process which redistributes the mass, changing the size of a collection of dust grains with a factor $C$.
The number density will then scale with $C^{-3}$, while the surface area of a single grain will scale as $C^{2}$.
The surface area provided by all the grains will hence scale as $C^{-1}$ \citep{1983ApJ...269L..57S}.
So shrinking the average dust grain size while conserving the mass or volume results in more surface area.}
As an example, \citet{2005MNRAS.362..666L} have shown that a broad size distribution of dust grains enhances the production of H$_2$.

{Besides the surface area effect, the efficiency of H$_2$ formation on certain types of dust grains could play a role.
The grain temperature affects various processes that determine the efficiency, such as the diffusion rate and evaporation time of atoms and molecules on the grain surface.
The review by \cite{2017MolAs...9....1W} explains the possible effects in depth, and section 4.1 of this review provides an overview of which processes are the most important for several regimes in the diffuse ISM.
For small grains in particular, the temperature undergoes random fluctuations, jumping between high and low temperatures as the grains absorb UV photons and emit in the IR.
This can allow small grains to have a high formation efficiency, even when the average temperature is high \citep{2014A&A...569A.100B}.}

{Small grains can also consist of large molecules with specific surface properties.
Considering that the FUV rise is the most important term in the \nhtwo\  correlation, the carriers of this feature could have surface properties that make them the dominant contributor to H$_2$ formation.
PAHs have been found to be an important H$_2$ formation pathway, at temperatures $\gtrsim \SI{200}{\kelvin}$ and in dense PDRs \citep{2009ApJ...704..274L, 2015A&A...579A..72B, 2018A&A...616A.167C}.
There are various theoretical and experimental results confirming that PAHs can be superhydrogenated when H atoms stick to them, and become catalyst for H$_2$ formation \citep{2008ApJ...679..531R, 2012ApJ...752....3T, 2012ApJ...745L...2M}.
Hydrogenated amorphous carbon grains could also play a role, through a photolysis process that forms H$_2$ \citep{2014A&A...569A.119A, 2015A&A...581A..92J}.
Both PAHs and small hydrogenated amorphous carbon grains have optical properties that make them possible candidates for the NUV bump and FUV rise \citep{1992ApJ...393L..79J, 2013A&A...558A..62J}.}

\subsubsection{Small Grains Survive Only in H$_2$-dominated Environment}
The second scenario is that the presence of molecular gas, or the environment in which it appears, promotes the abundance of {the small in grains question}.
This could be either because those shielded environments prevent photo-destruction of the grains, or because the density and temperature conditions are more favorable for the formation processes of the grains.

\citet{2018A&A...613A...9M} found that grains are larger in more strongly irradiated regions of the ISM, by comparing $R_{5495}$ (similar to $\rv$) in \ion{H}{2} regions and hot bubbles with molecular sightlines.
They suggest that selective destruction of small dust grains by EUV photons or thermal sputtering by atoms or ions causes increased values of $R_{5495}$ in hotter regions.
In dense molecular regions on the other hand, \rv\ has also been found to be higher due to coagulation causing a growth in big grains and a removal of small grains \citep{1993ApJ...407..806C, 2009A&A...502..845O}.
For example, in TMC-1, such processes have resulted in a extinction curve without \SI{2175}{\angstrom} bump, showing  that the small grains carrying this feature have been removed \citep{2004ApJ...602..291W}.
The fact that the average grain size as traced by \rv\ is larger in both hot bubbles and dense molecular clouds, means that the relative abundance of small grains must reach a peak somewhere between those two regimes.
Perhaps this is in the diffuse neutral regime which we are focusing on.

The plots including $T_{01}$ and \denh\ in Figure~\ref{fig:tempdens} showed us that the highest values of \cav{4} ($\gtrsim 0.2$) only occur for sightlines of low temperature ($T_{01} \lesssim 75 $ K) and higher density ($\nh / d \gtrsim \SI{1}{\per\cubic\cm}$) in this sample, {while the results of Section \ref{sec:column} showed that the FUV rise carrier coincides with molecular gas.}
This could mean that {the sub-population of small grains carrying the FUV feature}, preferably survives in colder, denser regions of the {diffuse molecular} ISM.
Similar behavior has been suggested for PAHs by {\citet{2022ApJ...929...23H}}, who found a correlation between the cold neutral medium fraction and the fraction of dust in PAHs, in an all-sky map of the cold neutral medium.
It is plausible that PAHs are {at least} partly responsible for {the UV extinction features}.
One indication is that IR emission features associated with PAHs correlate with the strength of the dust extinction bump in the NUV, between sightlines \citep{1993A&A...275..549J, 2017ApJ...836..173B, 2022ApJ...925...19M} 
and between averages over the MW, LMC and SMC \citep{2002A&A...382.1042V}.
{Additionally there are indications that volatile hydrocarbon dust is reponsible for at least part of the FUV extinction, linking the FUV extinction to the lifecycle of evaporation, condensation, and possible photo-destruction of such macromolecules \citep{1994ApJ...430..630B}.}
The PAH fraction has also been found to correlate with molecular gas in the SMC \citep{2010ApJ...715..701S}.
However, sightlines in the SMC typically have a very weak or non-existent bump, while some still have an FUV rise \cite{2003ApJ...594..279G}.
Therefore, if one assumes that PAHs always cause a substantial NUV bump, then there should exist another small grain component, which only contributes to the FUV rise.

\subsubsection{Simulation suggestions}
\label{sec:simu}

{To determine if scenario \#1 is viable, one would need to quantify how strongly the observed differences in the dust properties affect the H$_2$ fraction of the gas.
We suggest that PDR models could be used for this purpose, as long as they have a flexible enough dust model, one that can represent the variations in \rv\ and the FUV rise. 
Both the dust extinction affecting the radiative transfer in the model, and the models for physical processes such as H$_2$ formation on dust grain surfaces and heating by the photoelectric on dust grains, should depend on the same underlying grain size distribution.
If one finds that certain dust properties are required to form a substantial amount of H$_2$, within the constraint typical for our sample of $\av \lesssim 2$, then the simulations would support scenario \#1.
On the other hand, if these dust variations only cause a minor boost in the H$_2$ fraction, then the FUV rise vs.\ \nhtwo\ relation likely has another cause.}

{Scenario \#2 is harder to model, as this requires knowledge of the destruction processes of small grains, which need to be simulated simultaneously with H$_2$ formation and dissociation.
The galaxy evolution simulation by \citet{2022MNRAS.514.1461R} includes a variable grain size distribution together with H$_2$ formation and dissociation, and shows a jump in the molecular fraction as a function of their metric for the small-to-large grain ratio.
An initial step would be to calculate the properties along randomly selected sightlines of this model, and see if the same statistical properties are observed compared to our observations of MW sightlines.
In that way our results could provide constraints on this type of simulation.}

\subsection{Dust Evolution}
\label{sec:dustevolution}

We compared \nhav\ and \fhtwo\ to investigate if the phase of the medium plays a role in the variations of \nhav, but there is a null result for any correlation between \nhav\ and \fhtwo.
In other words, we find no evidence that dust grows preferentially in more molecular regions of the diffuse Milky Way ISM, at least not within our parameter space and when using \nhav\ as a tracer for the gas-to-dust ratio.
This is within expectations, as FIR studies (Clark et al. \textit{in prep}, \citealt{2017ApJ...841...72R}) and depletions \citep{2021ApJ...910...95R} suggest that the dust-to-gas ratio reaches a plateau at $\nh\ \geq \SI{1e21}{\per\square\cm}$, as most of the available metals in the gas have already been locked up in dust at that column density. 
Most of the dust-to-gas ratio variations happen at lower column densities, which we are not probing with our sightlines. 

The remaining variations in \nhav\ do correlate positively with the UV-to-optical extinction ratio however, according to our results shown in Figure~\ref{fig:rvtrend}. 
In other words, increases in the amount of dust absorbing in the V-band (relative to the gas) go along with relatively fewer small grains.
Especially notable are four low-\nhav\ sightlines, highlighted on the scatter plots using red squares.
They have a wide range of \fhtwo\ values, but similarly low \nhav, \rv, \ak, and \abumpav\ values.
These four sightlines seem to have undergone a more drastic form of dust growth, which has increased \av\ relative to the gas, while flattening the FUV rise, as seen from their highlighted position in the \cav{4} panel of in Figure~\ref{fig:fh2.vs.params}.
The value of \ak\ for these four sightlines is not an outlier of the \nhtwo\ relationship shown in Figure~\ref{fig:columns}.
Meanwhile, \ak/\av\ seems to be near the minimum of the range of values in the sample (Figure~\ref{fig:rvtrend}), and the same goes for the FUV bump and rise parameters (Figure~\ref{fig:fh2.vs.params}), which means those features become small relative to the total extinction.
More data points with \rvi\ between 0.2 and 0.3 could shed further light on whether the four low-\nhav\ sightlines are just anomalies, or if there is a global trend across the entire \rvi\ range.
Since one of these four sightlines is HD200775, a star behind the nebula NGC\,7023 which contains a PDR, the increased \rv\ observed for this sightline might be related to the changes in \rv\ observed in the IC\,63 PDR \citep{2020ApJ...888...22V}, where \rv\ increases near the dense part of the PDR front.

It seems unlikely that the changes in the UV-to-optical extinction ratio are driven by the destruction/survival balance of the small grains, as that would not affect \nhav\ because small grains have only a limited amount of absorption in the V-band.
According to theoretical and numerical models of dust evolution and extinction curves in galaxies \citep{2012MNRAS.422.1263H, 2013MNRAS.432..637A, 2017MNRAS.469..870H, 2019MNRAS.482.2555H, 2020MNRAS.491.3844A}, extinction curves are initially flat because stellar dust production results in predominantly large grains.
The FUV rise and bump become more prominent under the effects of shattering in the diffuse ISM, which produces small grains at the expense of large grains, and by the accretion of gas-phase metals in the dense ISM, which primarily enhances small grains.
Coagulation then depletes the small grain population as they stick to the larger grains, increasing the dust optical depth at longer wavelengths.
These simulations have been combined with observations of three nearby galaxies, to find that regions with high metallicity and molecular fraction are affected by both accretion and coagulation \citep{2020A&A...636A..18R}.
For our diffuse Milky Way sample, we observe that a decrease in \nhav\ (increase in big grains per gas unit) is coupled with a decrease in \akav\ and \abumpav\ (decrease in small grains), so these trends likely originate from the balance between coagulation and shattering effects.

\section{Conclusions}
\label{sec:conclusions}

We investigated the correlations between the \hi\ and H$_2$ gas contents and dust extinction in the diffuse Milky Way ISM, by combining the \citet{2009ApJ...705.1320G} extinction data with accurate measurements of the \hi\  and H$_2$ column densities.
Using distances from Gaia DR2, and H$_2$ excitation data, several derived quantities were calculated, most notably the average number density \denh\ and the rotational temperature $T_{01}$ of H$_2$.
Correlations between pairs of quantities were explored, while taking into account both the errors in x and y, and correlations between the measurement uncertainties of derived quantities that depend on a common factor.
The following notable trends were found:
\begin{enumerate}
\item The molecular column \nhtwo\ correlates very strongly with \ak, but not with \av, {indicating that a portion of the carriers of the FUV extinction coincide with H$_2$ along these diffuse sightlines}.
{This correlation is caused by a linear relationship between the contribution of the FUV rise term to \ak, showing that the carriers of the FUV rise feature appear primarily in molecular gas. 
We discussed two scenarios: either the FUV rise carriers are crucial for H$_2$ formation, or they coincide with H$_2$ because they require similar physical conditions to form and survive.}
\item The connection between small grains and H$_2$ and is further highlighted by a positive correlation between \fhtwo, and the FM90 parameters for the UV bump width $\gamma$ and the FUV rise strength \cav{4}.
\item The highest values of \cav{4} coincide with the lowest values for the estimated number density \denh, and the H$_2$ rotational temperature $T_{01}$.
This implies that the {FUV rise carriers} have a higher abundance in denser and colder regions of the diffuse molecular gas.
\item The total hydrogen column \nh\ shows a strong linear trend with the optical dust extinction measurements \av\ and \ebv, as expected from earlier literature.
Deviations from the main trend in \nhav\ are partially explained by a positive correlation with parameters that indicate the small-to-large grain ratio: \rvi, \akav, and \abumpav.
The effects of coagulation are a possible interpretation for the observed reduction in small grains and enhanced extinction at optical wavelengths.
\end{enumerate}

The connection between {the FUV rise carrier} and H$_2$, which we consider the main conclusion of this work, raises new questions. 
The first open question is which specific type or class of dust particles, {most likely carbonaceous}, shows such a strong preference for H$_2$-dominated environments.
To solve this question, different candidates that absorb mostly in the FUV could be considered, and their viability could be evaluated according to several criteria, including but not limited to: 1) They are destroyed in atomic or ionized regions 2) They survive in the diffuse molecular regime 3) Their optical depth, cross section, number density, and elemental composition need to be compatible with constraints based on depletions.

The second question is the exact mechanism through which the observed {FUV rise}-\nhtwo\ correlation occurs.
{As suggested in Section \ref{sec:simu}, we believe results from theoretical PDR models might be able to disentangle whether the extra extinction and H$_2$ formation by the FUV rise particles enhance the H$_2$ content, or if this grain population and H$_2$ simply trace the same physical conditions in the diffuse ISM.}

\begin{acknowledgments}
  This study was supported by NASA ADP grant NAG5-13033.
  Based on observations made with the NASA-CNES-CSA Far Ultraviolet Spectroscopic Explorer.
  FUSE is operated for NASA by the Johns Hopkins University under NASA contract NAS5-32985.
  The IUE and HST data used in this paper were obtained from the Mikulski Archive for Space Telescopes (MAST) at the Space Telescope Science Institute.
  The specific observations analyzed can be accessed via \dataset[https://doi.org/10.17909/5qq7-4c81]{https://doi.org/10.17909/5qq7-4c81}.
  STScI is operated by the Association of Universities for Research in Astronomy, Inc., under NASA contract NAS5–26555.
  Support to MAST for these data is provided by the NASA Office of Space Science via grant NAG5–7584 and by other grants and contracts.
  This research has made use of the SIMBAD database, operated at CDS (Strasbourg, France).  
\end{acknowledgments}

\software{
  The analysis and plotting code written specifically for this work is available on the GitHub profile of the author (\url{https://github.com/drvdputt/dust_fuse_h2}).
  Persistent version on Zenodo: \citet{dries_van_de_putte_2022_7108380}.
  Packages: astropy \citep{2013A&A...558A..33A, 2018AJ....156..123A}, scipy \citep{2020NatMe..17..261V}.
}


\bibliography{bibliography}

\end{document}

%% file: h2_edit.tex
\startlongtable
\begin{deluxetable*}{lr@{$\pm$}lr@{$\pm$}lr@{$\pm$}lr@{$\pm$}lr@{$\pm$}lr@{$\pm$}lr@{$\pm$}lr@{$\pm$}l}
  \tablecaption{Molecular hydrogen column density data}
  \tablewidth{0pt}
  \tablehead{
    \colhead{Star} &
    \multicolumn{2}{c}{$N({\rm J}=0)$} & \multicolumn{2}{c}{$N(1)$} &
    \multicolumn{2}{c}{$N(2)$} & \multicolumn{2}{c}{$N(3)$} &
    \multicolumn{2}{c}{$N(4)$} & \multicolumn{2}{c}{$N(5)$} &
    \multicolumn{2}{c}{$N(6)$} & \multicolumn{2}{c}{$N(7)$}
  }
  \startdata
  \multicolumn{17}{c}{Reddened Stars} \\
  BD$+$35 4258 & 19.10 & 0.10 & 19.35 & 0.10 & 17.45 & 0.35 & 17.25 & 0.40 &
  15.50 & 0.10 & 15.20 & 0.06 & \multicolumn{2}{c}{\nodata} &
  \multicolumn{2}{c}{\nodata} \\
  BD$+$53 2820 & 19.55 & 0.25 & 19.80 & 0.25 & 17.45 & 0.50 & 17.05 & 0.50 &
  15.25 & 0.25 & 14.65 & 0.15 & \multicolumn{2}{c}{\nodata} &
  \multicolumn{2}{c}{\nodata} \\
  & 19.00 & 0.50 & 19.00 & 0.50 & 15.40 & 0.25 & 15.40 & 0.25 &
  14.60 & 0.25 & 14.30 & 0.25 & \multicolumn{2}{c}{\nodata} &
  \multicolumn{2}{c}{\nodata} \\
  BD$+$56 524  & 19.65 & 0.40 & 20.40 & 0.40 & 18.55 & 0.25 & 16.45 & 0.25 &
  15.65 & 0.25 & 15.00 & 0.25 & \multicolumn{2}{c}{\nodata} &
  \multicolumn{2}{c}{\nodata} \\
  & 20.45 & 0.40 & 19.85 & 0.40 & 18.75 & 0.25 & 16.65 & 0.25 &
  14.85 & 0.25 & 14.65 & 0.25 & \multicolumn{2}{c}{\nodata} &
  \multicolumn{2}{c}{\nodata} \\
  HD001383     & 19.80 & 0.25 & 19.85 & 0.25 & 17.20 & 0.25 & 16.55 & 0.25 &
  15.10 & 0.25 & 14.75 & 0.25 & \multicolumn{2}{c}{\nodata} &
  \multicolumn{2}{c}{\nodata} \\
  & 19.60 & 0.25 & 19.85 & 0.25 & 17.00 & 0.25 & 16.50 & 0.25 &
  15.10 & 0.25 & 14.60 & 0.25 & \multicolumn{2}{c}{\nodata} &
  \multicolumn{2}{c}{\nodata} \\
  & 19.00 & 0.25 & 19.35 & 0.25 & 16.20 & 0.25 & 15.15 & 0.25 &
  13.90 & 0.25 & 13.60 & 0.25 & \multicolumn{2}{c}{\nodata} &
  \multicolumn{2}{c}{\nodata} \\
  HD013268     & 19.98 & 0.10 & 20.22 & 0.10 & 17.02 & 0.50 & 16.57 & 0.40 &
  15.67 & 0.10 & 15.19 & 0.08 & \multicolumn{2}{c}{\nodata} &
  \multicolumn{2}{c}{\nodata} \\
  HD014250     & 20.15 & 0.25 & 20.40 & 0.25 & 18.70 & 0.15 & 18.05 & 0.25 &
  15.85 & 0.15 & 14.90 & 0.15 & \multicolumn{2}{c}{\nodata} &
  \multicolumn{2}{c}{\nodata} \\
  & 19.55 & 0.50 & 19.40 & 0.25 & 16.00 & 0.50 & 16.00 & 0.25 &
  15.05 & 0.15 & 14.70 & 0.15 & \multicolumn{2}{c}{\nodata} &
  \multicolumn{2}{c}{\nodata} \\
  HD014434     & 20.10 & 0.25 & 20.30 & 0.15 & 18.00 & 0.50 & 16.90 & 0.15 &
  15.40 & 0.10 & 14.80 & 0.15 & \multicolumn{2}{c}{\nodata} &
  \multicolumn{2}{c}{\nodata} \\
  & 18.80 & 0.25 & 19.00 & 0.25 & 17.80 & 0.50 & 16.80 & 0.25 &
  15.30 & 0.15 & 14.85 & 0.15 & \multicolumn{2}{c}{\nodata} &
  \multicolumn{2}{c}{\nodata} \\
  HD015558     & 20.39 & 0.09 & 20.58 & 0.09 & 18.15 & 0.65 & 16.96 & 1.17 &
  15.70 & 0.52 & 15.26 & 0.22 & \multicolumn{2}{c}{\nodata} &
  \multicolumn{2}{c}{\nodata } \\
  HD017505     & 20.35 & 0.10 & 20.59 & 0.10 & 18.04 & 0.40 & 17.20 & 0.60 &
  15.82 & 0.20 & 15.47 & 0.20 & 14.68 & 0.13 & 14.40 & 0.13 \\
  HD023060     & 20.34 & 0.09 & 20.13 & 0.09 & 18.42 & 0.05 & 17.10 & 0.40 &
  15.58 & 0.25 & 14.74 & 0.15 & \multicolumn{2}{c}{\nodata} &
  \multicolumn{2}{c}{\nodata} \\
  HD027778\tablenotemark{a}     & 20.60 & 0.10 & 20.10 & 0.10 & 18.54 & 0.08 & 17.47 & 0.17 &
  15.32 & 0.42 & 14.69 & 0.15 & \multicolumn{2}{c}{\nodata} &
  \multicolumn{2}{c}{\nodata} \\
  HD037903\tablenotemark{a}     & 20.54 & 0.09 & 20.50 & 0.12 & 19.60 & 0.09 & 18.56 & 0.07 &
  17.08 & 0.46 & 15.61 & 0.11 & 14.99 & 0.35 & 14.49 & 0.30 \\
  HD038087\tablenotemark{a}     & 20.37 & 0.09 & 20.30 & 0.09 & 19.10 & 0.13 & 18.21 & 0.10 &
  16.41 & 0.65 & 15.84 & 0.50 & 14.89 & 0.30 & 14.61 & 0.20 \\
  HD045314     & 20.23 & 0.07 & 20.25 & 0.09 & 18.19 & 0.12 & 17.32 & 0.34 &
  15.12 & 0.17 & 14.34 & 0.05 & \multicolumn{2}{c}{\nodata} &
  \multicolumn{2}{c}{\nodata} \\
  HD046056\tablenotemark{a}     & 20.44 & 0.10 & 20.38 & 0.10 & 18.29 & 0.10 & 17.37 & 0.30 &
  15.64 & 0.17 & 15.02 & 0.13 & \multicolumn{2}{c}{\nodata} &
  \multicolumn{2}{c}{\nodata} \\
  HD046150     & 20.24 & 0.08 & 20.24 & 0.12 & 17.67 & 0.40 & 16.90 & 0.50 &
  15.41 & 0.09 & 14.90 & 0.05 & \multicolumn{2}{c}{\nodata} &
  \multicolumn{2}{c}{\nodata} \\
  HD046202\tablenotemark{a}     & 20.36 & 0.09 & 20.29 & 0.09 & 18.20 & 0.22 & 17.54 & 0.37 &
  15.51 & 0.24 & 15.10 & 0.14 & \multicolumn{2}{c}{\nodata} &
  \multicolumn{2}{c}{\nodata} \\
  HD047129     & 20.01 & 0.09 & 20.14 & 0.09 & 17.99 & 0.17 & 17.11 & 0.36 &
  15.51 & 0.13 & 14.90 & 0.05 & \multicolumn{2}{c}{\nodata} &
  \multicolumn{2}{c}{\nodata} \\
  HD047240     & 20.03 & 0.12 & 20.06 & 0.12 & 18.14 & 0.07 & 17.75 & 0.11 &
  15.83 & 0.14 & 15.60 & 0.12 & 14.51 & 0.09 & 14.43 & 0.10 \\
  HD047417     & 20.20 & 0.08 & 20.10 & 0.15 & 17.85 & 0.22 & 17.00 & 0.43 &
  15.30 & 0.06 & 14.70 & 0.05 & \multicolumn{2}{c}{\nodata} &
  \multicolumn{2}{c}{\nodata} \\
  HD062542\tablenotemark{a}     & 20.75 & 0.25 & 20.13 & 0.25 & 19.01 & 0.25 & 16.29 & 0.25 &
  15.69 & 0.25 & \multicolumn{2}{c}{\nodata} & \multicolumn{2}{c}{\nodata} &
  \multicolumn{2}{c}{\nodata} \\
  HD073882\tablenotemark{a}     & 21.00 & 0.09 & 20.45 & 0.09 & 18.87 & 0.05 & 18.05 & 0.11 &
  15.81 & 0.09 & 15.43 & 0.06 & \multicolumn{2}{c}{\nodata} &
  \multicolumn{2}{c}{\nodata} \\
  HD091651     & 18.75 & 0.15 & 18.80 & 0.15 & 17.00 & 0.42 & 17.00 & 0.40 &
  15.45 & 0.08 & 15.05 & 0.08 & 14.30 & 0.25 & 14.00 & 0.50 \\
  HD093222     & 19.50 & 0.10 & 19.40 & 0.15 & 18.20 & 0.25 & 17.75 & 0.30 &
  15.10 & 0.15 & 14.40 & 0.15 & \multicolumn{2}{c}{\nodata} &
  \multicolumn{2}{c}{\nodata} \\
  HD093250     & 19.90 & 0.25 & 20.12 & 0.15 & 17.35 & 0.50 & 17.38 & 0.40 &
  16.10 & 0.20 & 16.06 & 0.30 & 15.30 & 0.30 & 15.40 & 0.30 \\
  HD096675\tablenotemark{a}     & 20.63 & 0.15 & 20.35 & 0.10 & 18.55 & 0.07 & 16.85 & 0.38 &
  15.45 & 0.14 & 14.50 & 0.17 & \multicolumn{2}{c}{\nodata} &
  \multicolumn{2}{c}{\nodata} \\
  HD096715     & 20.48 & 0.08 & 20.15 & 0.10 & 17.55 & 0.46 & 16.75 & 0.62 &
  15.55 & 0.18 & 14.80 & 0.16 & \multicolumn{2}{c}{\nodata} &
  \multicolumn{2}{c}{\nodata} \\
  HD099872     & 20.31 & 0.15 & 20.18 & 0.15 & 18.48 & 0.20 & 16.05 & 1.20 &
  14.90 & 0.85 & 14.42 & 0.80 & \multicolumn{2}{c}{\nodata} &
  \multicolumn{2}{c}{\nodata} \\
  HD099890     & 19.15 & 0.15 & 19.30 & 0.10 & 16.65 & 0.60 & 16.40 & 0.55 &
  14.93 & 0.09 & 14.45 & 0.06 & \multicolumn{2}{c}{\nodata} &
  \multicolumn{2}{c}{\nodata} \\
  HD100213     & 20.15 & 0.12 & 20.15 & 0.12 & 16.85 & 1.60 & 15.80 & 2.10 &
  14.90 & 0.60 & 14.45 & 0.25 & \multicolumn{2}{c}{\nodata} &
  \multicolumn{2}{c}{\nodata} \\
  HD101190     & 20.22 & 0.10 & 20.02 & 0.10 & 18.23 & 0.65 & 16.20 & 0.51 &
  14.72 & 0.08 & 14.29 & 0.08 & \multicolumn{2}{c}{\nodata} &
  \multicolumn{2}{c}{\nodata} \\
  HD101205     & 20.00 & 0.23 & 19.90 & 0.12 & 18.30 & 0.20 & 17.70 & 0.42 &
  15.13 & 0.46 & 14.76 & 0.25 & \multicolumn{2}{c}{\nodata} &
  \multicolumn{2}{c}{\nodata} \\
  HD103779     & 19.35 & 0.25 & 19.70 & 0.10 & 16.15 & 0.55 & 15.80 & 0.35 &
  14.60 & 0.15 & 14.30 & 0.18 & \multicolumn{2}{c}{\nodata} &
  \multicolumn{2}{c}{\nodata} \\
  HD122879     & 19.90 & 0.10 & 20.10 & 0.10 & 17.38 & 0.15 & 16.83 & 0.45 &
  15.30 & 0.40 & 14.70 & 0.25 & \multicolumn{2}{c}{\nodata} &
  \multicolumn{2}{c}{\nodata} \\
  HD124979     & 20.08 & 0.10 & 20.18 & 0.10 & 17.85 & 0.20 & 16.91 & 0.45 &
  15.38 & 0.10 & 14.70 & 0.10 & \multicolumn{2}{c}{\nodata} &
  \multicolumn{2}{c}{\nodata} \\
  HD147888\tablenotemark{a}     & 20.37 & 0.09 & 19.76 & 0.12 & 18.58 & 0.10 & 17.06 & 0.25 &
  15.70 & 0.44 & 15.24 & 0.38 & 14.38 & 0.15 & \multicolumn{2}{c}{\nodata} \\
  HD148422     & 19.85 & 0.10 & 19.80 & 0.10 & 17.55 & 0.30 & 16.85 & 0.30 &
  15.35 & 0.25 & 14.75 & 0.10 & \multicolumn{2}{c}{\nodata} &
  \multicolumn{2}{c}{\nodata} \\
  & 15.30 & 0.30 & 15.50 & 0.40 & 14.85 & 0.15 & 14.80 & 0.15 &
  13.75 & 0.25 & 14.30 & 0.10 & \multicolumn{2}{c}{\nodata} &
  \multicolumn{2}{c}{\nodata} \\
  HD149404\tablenotemark{a}     & 20.60 & 0.10 & 20.35 & 0.10 & 18.50 & 0.40 & 16.84 & 0.15 &
  15.94 & 0.15 & 15.34 & 0.10 & 14.00 & 0.25 & \multicolumn{2}{c}{\nodata} \\
  HD151805     & 20.10 & 0.10 & 20.01 & 0.08 & 18.22 & 0.18 & 16.90 & 0.55 &
  15.63 & 0.37 & 15.06 & 0.10 & \multicolumn{2}{c}{\nodata} &
  \multicolumn{2}{c}{\nodata} \\
  HD152233     & 20.05 & 0.15 & 19.95 & 0.15 & 18.40 & 0.10 & 17.55 & 0.25 &
  16.50 & 0.25 & 15.50 & 0.10 & \multicolumn{2}{c}{\nodata} &
  \multicolumn{2}{c}{\nodata} \\
  & 17.30 & 1.50 & 16.70 & 1.00 & 15.00 & 0.30 & 15.20 & 0.15 &
  14.65 & 0.25 & 14.10 & 0.15 & \multicolumn{2}{c}{\nodata} &
  \multicolumn{2}{c}{\nodata} \\
  HD152234     & 20.30 & 0.10 & 20.07 & 0.10 & 18.25 & 0.20 & 17.60 & 0.25 &
  15.90 & 0.15 & 15.27 & 0.25 & \multicolumn{2}{c}{\nodata} &
  \multicolumn{2}{c}{\nodata} \\
  HD152248     & 20.00 & 0.10 & 19.90 & 0.10 & 18.30 & 0.15 & 17.90 & 0.25 &
  15.70 & 0.25 & 15.00 & 0.25 & \multicolumn{2}{c}{\nodata} &
  \multicolumn{2}{c}{\nodata} \\
  HD152249     & 19.95 & 0.45 & 19.70 & 0.15 & 18.00 & 0.10 & 16.70 & 0.15 &
  14.85 & 0.12 & 14.00 & 0.25 & \multicolumn{2}{c}{\nodata} &
  \multicolumn{2}{c}{\nodata} \\
  & 19.45 & 0.25 & 19.65 & 0.15 & 17.10 & 0.10 & 16.65 & 0.10 &
  15.45 & 0.12 & 14.65 & 0.10 & \multicolumn{2}{c}{\nodata} &
  \multicolumn{2}{c}{\nodata} \\
  & 18.60 & 1.00 & 18.60 & 1.00 & 14.90 & 0.10 & 14.90 & 0.10 &
  14.25 & 0.10 & 14.05 & 0.10 & \multicolumn{2}{c}{\nodata} &
  \multicolumn{2}{c}{\nodata} \\
  HD152723     & 20.00 & 0.15 & 20.00 & 0.10 & 16.35 & 0.30 & 16.00 & 0.60 &
  15.10 & 0.25 & 14.80 & 0.10 & \multicolumn{2}{c}{\nodata} &
  \multicolumn{2}{c}{\nodata} \\
  HD157857     & 20.30 & 0.10 & 20.40 & 0.10 & 18.00 & 0.20 & 16.60 & 0.20 &
  15.05 & 0.10 & 14.65 & 0.10 & \multicolumn{2}{c}{\nodata} &
  \multicolumn{2}{c}{\nodata} \\
  HD160993     & 19.14 & 0.06 & 19.20 & 0.06 & 17.90 & 0.05 & 17.68 & 0.05 &
  15.63 & 0.09 & 14.74 & 0.07 & \multicolumn{2}{c}{\nodata} &
  \multicolumn{2}{c}{\nodata} \\
  HD163522     & 19.20 & 0.20 & 19.25 & 0.20 & 17.60 & 0.60 & 16.20 & 0.60 &
  14.80 & 0.10 & 14.25 & 0.10 & \multicolumn{2}{c}{\nodata} &
  \multicolumn{2}{c}{\nodata} \\
  & 18.20 & 0.50 & 18.50 & 0.20 & 15.10 & 0.20 & 15.00 & 0.10 &
  14.55 & 0.10 & 13.75 & 0.10 & \multicolumn{2}{c}{\nodata} &
  \multicolumn{2}{c}{\nodata} \\
  HD164816     & 19.66 & 0.06 & 19.50 & 0.08 & 17.56 & 0.10 & 17.04 & 0.25 &
  15.08 & 0.50 & 14.32 & 0.10 & \multicolumn{2}{c}{\nodata} &
  \multicolumn{2}{c}{\nodata} \\
  HD164906     & 19.98 & 0.10 & 19.89 & 0.10 & 18.30 & 0.10 & 17.15 & 0.35 &
  15.05 & 0.10 & 14.30 & 0.30 & \multicolumn{2}{c}{\nodata} &
  \multicolumn{2}{c}{\nodata} \\
  HD165052     & 19.75 & 0.15 & 19.95 & 0.15 & 17.35 & 0.60 & 16.80 & 0.20 &
  14.95 & 0.10 & 14.50 & 0.10 & \multicolumn{2}{c}{\nodata} &
  \multicolumn{2}{c}{\nodata} \\
  HD167402     & 19.95 & 0.10 & 19.75 & 0.15 & 18.20 & 0.10 & 17.85 & 0.40 &
  15.10 & 0.10 & 14.35 & 0.15 & \multicolumn{2}{c}{\nodata} &
  \multicolumn{2}{c}{\nodata} \\
  HD167771     & 20.40 & 0.15 & 20.35 & 0.10 & 18.27 & 0.25 & 16.12 & 0.15 &
  15.20 & 0.10 & 14.65 & 0.10 & \multicolumn{2}{c}{\nodata} &
  \multicolumn{2}{c}{\nodata} \\
  HD168076\tablenotemark{a}     & 20.45 & 0.10 & 20.30 & 0.10 & 18.30 & 0.30 & 17.20 & 0.50 &
  16.40 & 0.30 & 15.85 & 0.15 & 14.75 & 0.15 & 14.60 & 0.15 \\
  HD168941     & 19.98 & 0.10 & 19.82 & 0.10 & 17.76 & 0.50 & 17.20 & 0.15 &
  15.19 & 0.10 & 14.56 & 0.10 & \multicolumn{2}{c}{\nodata} &
  \multicolumn{2}{c}{\nodata} \\
  HD178487     & 20.24 & 0.10 & 20.15 & 0.15 & 18.21 & 0.08 & 17.83 & 0.13 &
  15.67 & 0.10 & 14.85 & 0.11 & \multicolumn{2}{c}{\nodata} &
  \multicolumn{2}{c}{\nodata} \\
  HD179406\tablenotemark{a}     & 20.46 & 0.08 & 20.25 & 0.08 & 18.10 & 0.15 & 17.05 & 0.30 &
  15.20 & 0.10 & 14.45 & 0.10 & \multicolumn{2}{c}{\nodata} &
  \multicolumn{2}{c}{\nodata} \\
  HD179407     & 20.03 & 0.10 & 19.99 & 0.10 & 17.80 & 0.32 & 17.01 & 0.66 &
  15.37 & 0.21 & 14.78 & 0.10 & \multicolumn{2}{c}{\nodata} &
  \multicolumn{2}{c}{\nodata} \\
  HD185418\tablenotemark{a}     & 20.29 & 0.09 & 20.48 & 0.09 & 18.32 & 0.05 & 17.34 & 0.13 &
  15.35 & 0.08 & 14.53 & 0.50 & \multicolumn{2}{c}{\nodata} &
  \multicolumn{2}{c}{\nodata} \\
  HD188001     & 19.77 & 0.10 & 19.97 & 0.10 & 17.89 & 0.18 & 16.89 & 0.70 &
  15.24 & 0.54 & 14.78 & 0.30 & \multicolumn{2}{c}{\nodata} &
  \multicolumn{2}{c}{\nodata} \\
  HD190603     & 20.35 & 0.10 & 20.40 & 0.10 & 18.40 & 0.10 & 17.58 & 0.12 &
  15.88 & 0.10 & 15.41 & 0.10 & \multicolumn{2}{c}{\nodata} &
  \multicolumn{2}{c}{\nodata} \\
  HD192639\tablenotemark{a}     & 20.33 & 0.10 & 20.45 & 0.10 & 18.41 & 0.09 & 17.51 & 0.17 &
  15.77 & 0.07 & 15.24 & 0.07 & 14.44 & 0.17 & \multicolumn{2}{c}{\nodata} \\
  HD197770     & 20.60 & 0.20 & 20.59 & 0.10 & 18.44 & 0.10 & 17.58 & 0.17 &
  16.20 & 0.71 & 15.15 & 0.60 & \multicolumn{2}{c}{\nodata} &
  \multicolumn{2}{c}{\nodata} \\
  HD198781     & 20.23 & 0.09 & 20.11 & 0.09 & 17.55 & 0.73 & 16.30 & 1.40 &
  14.90 & 0.11 & 14.35 & 0.14 & \multicolumn{2}{c}{\nodata} &
  \multicolumn{2}{c}{\nodata} \\
  HD199579\tablenotemark{a}     & 20.25 & 0.10 & 20.20 & 0.15 & 18.45 & 0.15 & 17.10 & 0.10 &
  15.95 & 0.05 & 15.55 & 0.05 & 14.30 & 0.05 & 14.35 & 0.05 \\
  HD200775     & 20.80 & 0.10 & 20.60 & 0.20 & 19.50 & 0.15 & 18.00 & 0.30 &
  17.05 & 0.15 & 15.50 & 1.30 & 14.90 & 1.20 & \multicolumn{2}{c}{\nodata} \\
  HD203938\tablenotemark{a}     & 20.75 & 0.09 & 20.65 & 0.09 & 19.10 & 0.10 & 17.71 & 0.20 &
  16.16 & 0.35 & 15.46 & 0.13 & \multicolumn{2}{c}{\nodata} &
  \multicolumn{2}{c}{\nodata} \\
  HD206267\tablenotemark{a}     & 20.62 & 0.06 & 20.35 & 0.09 & 17.94 & 0.21 & 16.76 & 0.58 &
  15.26 & 0.15 & 14.92 & 0.08 & \multicolumn{2}{c}{\nodata} &
  \multicolumn{2}{c}{\nodata} \\
  HD206773     & 20.00 & 0.15 & 20.20 & 0.10 & 18.20 & 0.08 & 17.40 & 0.18 &
  15.40 & 0.06 & 14.75 & 0.05 & \multicolumn{2}{c}{\nodata} &
  \multicolumn{2}{c}{\nodata} \\
  HD207198\tablenotemark{a}     & 20.60 & 0.10 & 20.45 & 0.10 & 18.40 & 0.10 & 17.00 & 0.15 &
  15.70 & 0.10 & 14.90 & 0.05 & \multicolumn{2}{c}{\nodata} &
  \multicolumn{2}{c}{\nodata} \\
  HD209339     & 19.85 & 0.10 & 20.00 & 0.10 & 18.00 & 0.18 & 17.08 & 0.24 &
  15.31 & 0.10 & 14.70 & 0.05 & \multicolumn{2}{c}{\nodata} &
  \multicolumn{2}{c}{\nodata} \\
  HD216898     & 20.60 & 0.10 & 20.80 & 0.10 & 18.82 & 0.14 & 17.60 & 0.34 & 
  15.97 & 0.10 & 15.36 & 0.15 & \multicolumn{2}{c}{\nodata} &
  \multicolumn{2}{c}{\nodata} \\
  HD239729     & 20.92 & 0.15 & 20.52 & 0.15 & 18.72 & 0.25 & 17.74 & 0.30 &
  15.36 & 1.00 & \multicolumn{2}{c}{\nodata} & \multicolumn{2}{c}{\nodata} &
  \multicolumn{2}{c}{\nodata} \\
  HD326329     & 19.98 & 0.06 & 19.96 & 0.06 & 18.20 & 0.25 & 17.70 & 0.35 & 
  15.70 & 0.10 & 15.15 & 0.12 & 13.70 & 0.60 & \multicolumn{2}{c}{\nodata} \\
  HD332407     & 20.04 & 0.10 & 20.15 & 0.10 & 17.76 & 0.25 & 16.87 & 0.35 & 
  15.61 & 0.06 & 15.16 & 0.07 & \multicolumn{2}{c}{\nodata} &
  \multicolumn{2}{c}{\nodata} \\
  \multicolumn{17}{c}{Comparison stars} \\
  BD+32 270    & 18.28 & 0.08 & 18.47 & 0.05 & 16.48 & 0.57 & 15.82 & 0.20 &
  14.60 & 0.10 & 13.73 & 0.10 & \multicolumn{2}{c}{\nodata} &
  \multicolumn{2}{c}{\nodata} \\
  BD+52 3210   & 19.40 & 0.09 & 19.71 & 0.09 & 16.75 & 0.25 & 16.75 & 0.25 &
  15.65 & 0.10 & 15.30 & 0.10 & \multicolumn{2}{c}{\nodata} &
  \multicolumn{2}{c}{\nodata} \\
  HD037332     & 15.50 & 0.25 & 15.08 & 0.25 & 14.58 & 0.25 & 14.30 & 0.25 &
  \multicolumn{2}{c}{\nodata} & \multicolumn{2}{c}{\nodata} &
  \multicolumn{2}{c}{\nodata} & \multicolumn{2}{c}{\nodata} \\
  HD037525     & 15.40 & 1.40 & 16.10 & 1.20 & 15.20 & 1.65 & 14.90 & 2.80 &
  \multicolumn{2}{c}{\nodata} & \multicolumn{2}{c}{\nodata} &
  \multicolumn{2}{c}{\nodata} & \multicolumn{2}{c}{\nodata} \\
  HD051013     & 13.40 & 0.15 & 14.04 & 0.15 & 13.60 & 0.15 & 13.90 & 0.15 &
  \multicolumn{2}{c}{\nodata} & \multicolumn{2}{c}{\nodata} &
  \multicolumn{2}{c}{\nodata} & \multicolumn{2}{c}{\nodata} \\
  HD075309     & 19.74 & 0.09 & 19.87 & 0.09 & 17.63 & 0.12 & 17.14 & 0.26 &
  15.18 & 0.12 & 14.76 & 0.07 & \multicolumn{2}{c}{\nodata} &
  \multicolumn{2}{c}{\nodata} \\
  HD091824     & 19.51 & 0.09 & 19.36 & 0.09 & 17.80 & 0.11 & 17.41 & 0.18 &
  15.25 & 0.05 & 14.72 & 0.05 & \multicolumn{2}{c}{\nodata} &
  \multicolumn{2}{c}{\nodata} \\
  HD091983     & 19.97 & 0.12 & 19.72 & 0.09 & 17.36 & 0.50 & 16.42 & 0.45 &
  15.00 & 0.20 & 14.67 & 0.15 & \multicolumn{2}{c}{\nodata} &
  \multicolumn{2}{c}{\nodata} \\
  HD093028     & 19.10 & 0.12 & 19.26 & 0.12 & 17.77 & 0.09 & 17.52 & 0.13 &
  14.92 & 0.11 & 14.49 & 0.05 & \multicolumn{2}{c}{\nodata} &
  \multicolumn{2}{c}{\nodata} \\
  HD094493     & 19.85 & 0.12 & 19.80 & 0.09 & 18.04 & 0.14 & 17.45 & 0.26 &
  15.15 & 0.07 & 14.66 & 0.05 & \multicolumn{2}{c}{\nodata} &
  \multicolumn{2}{c}{\nodata} \\
  HD097471     & 19.65 & 0.10 & 19.50 & 0.10 & 17.58 & 0.34 & 17.14 & 0.43 &
  15.41 & 0.11 & 14.85 & 0.05 & \multicolumn{2}{c}{\nodata} &
  \multicolumn{2}{c}{\nodata} \\
  HD100276     & 19.27 & 0.15 & 19.26 & 0.10 & 17.56 & 0.60 & 16.73 & 0.60 &
  15.23 & 0.40 & 14.57 & 0.10 & \multicolumn{2}{c}{\nodata} &
  \multicolumn{2}{c}{\nodata} \\
  & 19.07 & 0.10 & 19.27 & 0.10 & 16.23 & 0.15 & 15.72 & 0.60 &
  14.65 & 0.25 & 14.17 & 0.10 & \multicolumn{2}{c}{\nodata} &
  \multicolumn{2}{c}{\nodata} \\
  HD104705     & 19.64 & 0.10 & 19.66 & 0.15 & 15.95 & 0.20 & 15.85 & 0.20 &
  15.00 & 0.20 & 14.38 & 0.10 & \multicolumn{2}{c}{\nodata} &
  \multicolumn{2}{c}{\nodata} \\
  & 16.48 & 1.00 & 16.10 & 1.00 & 15.95 & 0.20 & 15.27 & 0.20 &
  14.10 & 0.30 & 14.00 & 0.20 & \multicolumn{2}{c}{\nodata} &
  \multicolumn{2}{c}{\nodata} \\
  HD114444     & 18.20 & 0.25 & 18.10 & 0.25 & 16.00 & 0.35 & 15.50 & 0.50 &
  14.30 & 0.20 & \multicolumn{2}{c}{\nodata} & \multicolumn{2}{c}{\nodata} &
  \multicolumn{2}{c}{\nodata} \\
  & 19.40 & 0.15 & 19.60 & 0.15 & 16.10 & 0.35 & 15.60 & 0.50 &
  14.50 & 0.10 & \multicolumn{2}{c}{\nodata} & \multicolumn{2}{c}{\nodata} &
  \multicolumn{2}{c}{\nodata} \\
  HD116852     & 19.54 & 0.10 & 19.45 & 0.10 & 17.61 & 0.05 & 17.37 & 0.06 &
  14.97 & 0.05 & 14.11 & 0.05 & \multicolumn{2}{c}{\nodata} &
  \multicolumn{2}{c}{\nodata} \\
  HD172140     & 19.00 & 0.15 & 18.80 & 0.20 & 16.23 & 0.10 & 15.82 & 0.10 &
  14.44 & 0.10 & 14.01 & 0.10 & \multicolumn{2}{c}{\nodata} &
  \multicolumn{2}{c}{\nodata} \\
  HD210809     & 18.15 & 0.75 & 18.50 & 0.50 & 15.40 & 0.20 & 16.00 & 0.40 &
  14.60 & 0.20 & 14.15 & 0.15 & \multicolumn{2}{c}{\nodata} &
  \multicolumn{2}{c}{\nodata} \\
  & 19.00 & 0.35 & 19.45 & 0.20 & 15.25 & 0.30 & 14.70 & 0.30 &
  15.70 & 0.15 & 15.05 & 0.10 & \multicolumn{2}{c}{\nodata} &
  \multicolumn{2}{c}{\nodata} \\
  & 19.40 & 0.35 & 18.65 & 0.35 & 17.25 & 1.00 & 17.35 & 0.50 &
  14.15 & 0.15 & 13.70 & 0.25 & \multicolumn{2}{c}{\nodata} &
  \multicolumn{2}{c}{\nodata} \\
  & 18.10 & 0.75 & 19.00 & 0.35 & 15.30 & 0.20 & 15.00 & 0.20 &
  14.75 & 0.10 & 14.40 & 0.10 & \multicolumn{2}{c}{\nodata} &
  \multicolumn{2}{c}{\nodata} \\
  HD235874     & 19.20 & 0.10 & 19.45 & 0.10 & 17.70 & 0.25 & 17.40 & 0.25 &
  15.72 & 0.25 & 15.20 & 0.25 & \multicolumn{2}{c}{\nodata} &
  \multicolumn{2}{c}{\nodata} \\
  \multicolumn{17}{c}{Extra (no extinction curve)} \\
  HD003827   & 16.89 & 0.15 & 17.47 & 0.15 & 16.44 & 0.15 & 16.07 & 0.15 &
  14.57 & 0.08 & 14.17 & 0.08 & \multicolumn{2}{c}{\nodata} &
  \multicolumn{2}{c}{\nodata} \\
  HD022586   & 14.92 & 0.10 & 15.20 & 0.10 & 14.62 & 0.20 & 14.35 & 0.15 &
  \multicolumn{2}{c}{\nodata} & \multicolumn{2}{c}{\nodata} &
  \multicolumn{2}{c}{\nodata} & \multicolumn{2}{c}{\nodata} \\
  HD177989   & 20.00 & 0.10 & 19.65 & 0.10 & 18.00 & 0.30 & 17.05 & 0.20 &
  15.05 & 0.10 & 14.47 & 0.10 & \multicolumn{2}{c}{\nodata} &
  \multicolumn{2}{c}{\nodata} \\
  HD210121\tablenotemark{a}     & 20.63 & 0.15 & 20.13 & 0.15 & 18.71 & 0.30 & 18.10 & 0.30 &
  16.50 & 0.90 & 14.80 & 0.75 & \multicolumn{2}{c}{\nodata} &
  \multicolumn{2}{c}{\nodata} \\
  HD210839\tablenotemark{a}     & 20.50 & 0.10 & 20.40 & 0.10 & 18.40 & 0.05 & 17.20 & 0.15 &
  16.20 & 0.10 & 15.70 & 0.10 & 14.50 & 0.05 & 14.20 & 0.05 \\
  HD220057     & 20.05 & 0.10 & 19.90 & 0.10 & 17.70 & 0.10 & 16.10 & 0.20 &
  14.85 & 0.10 & 14.35 & 0.15 & \multicolumn{2}{c}{\nodata} &
  \multicolumn{2}{c}{\nodata} \\
  HD239683     & 20.42 & 0.30 & 20.48 & 0.25 & 18.69 & 0.25 & 17.75 & 0.25 &
  15.66 & 0.25 & 14.88 & 0.25 & \multicolumn{2}{c}{\nodata} &
  \multicolumn{2}{c}{\nodata} \\
  HD303308     & 15.00 & 0.50 & 15.00 & 0.50 & 15.00 & 0.50 & 14.40 & 0.50 &
  13.60 & 0.50 & 13.30 & 0.50 & 13.00 & 0.50 & 12.00 & 0.50 \\
  & 16.80 & 0.50 & 17.00 & 0.50 & 15.80 & 0.50 & 15.50 & 0.50 &
  16.00 & 0.50 & 15.40 & 0.50 & 15.40 & 0.50 & 15.00 & 0.50 \\
  & 19.20 & 0.25 & 19.35 & 0.25 & 17.50 & 0.50 & 16.80 & 0.50 &
  16.00 & 0.50 & 15.90 & 0.50 & 15.50 & 0.25 & 15.60 & 0.50 \\
  & 19.25 & 0.25 & 19.50 & 0.25 & 17.40 & 0.50 & 16.30 & 0.50 &
  16.00 & 0.50 & 15.90 & 0.25 & 14.80 & 0.25 & 14.70 & 0.25 \\
  & 19.35 & 0.25 & 19.60 & 0.25 & 17.20 & 0.50 & 15.60 & 0.50 &
  16.00 & 0.50 & 15.30 & 0.25 & 14.80 & 0.50 & 14.60 & 0.50 \\
  & 17.00 & 0.50 & 16.00 & 0.50 & 14.50 & 0.50 & 14.50 & 0.50 &
  13.50 & 0.50 & 13.70 & 0.50 & 13.00 & 0.00 & 13.00 & 0.50 \\
  HD315021     & 19.55 & 0.15 & 19.75 & 0.15 & 17.60 & 0.20 & 16.10 & 0.50 &
  14.45 & 0.10 & 14.00 & 0.15 & \multicolumn{2}{c}{\nodata} &
  \multicolumn{2}{c}{\nodata} \\
  & 18.60 & 0.20 & 18.65 & 0.20 & 15.25 & 0.20 & 14.90 & 0.10 &
  13.72 & 0.20 & 13.80 & 0.10 & \multicolumn{2}{c}{\nodata} &
  \multicolumn{2}{c}{\nodata} \\
  \enddata
  \tablenotetext{a}{FUSE H$_2$ also measured by \citet{2002ApJ...577..221R,2009ApJS..180..125R}}
  \tablecomments{Table is divided into three sections, with 75 reddened stars and 18 comparison stars of the GCC09 sample, as well as 9 extra sightlines for which extinction curves could not be derived by GCC09. All column densities are expressed in logarithmic form for absolute values in cm$^{-2}$. For models with multiple velocity components, one data line is given per component. \label{tab:h2dat}}

\end{deluxetable*}

%% file: hi_edit.tex
\startlongtable\begin{deluxetable*}{lr@{$\pm$}lr@{$\pm$}lr@{$\pm$}lr@{$\pm$}lccc}
\tablecaption{Total gas column densities and additional quantities \label{tab:gasdetails}}
\tablewidth{0pt}
\tablehead{
    \colhead{Star}
 & \multicolumn{2}{c}{$\log_{10} N(\text{H})$}
 & \multicolumn{2}{c}{$\log_{10} N(\text{\ion{H}{1}})$}
 & \multicolumn{2}{c}{$\log_{10} N(\text{H}_2)$}
 & \multicolumn{2}{c}{$d$}
 & \colhead{$f(\text{H}_2)$}
 & \colhead{$\log_{10} n(\text{H})$}
 & \colhead{HI Ref.} \\ 
 \colhead{}
 & \multicolumn{2}{c}{(\si{\per\square\cm})}
 & \multicolumn{2}{c}{(\si{\per\square\cm})}
 & \multicolumn{2}{c}{(\si{\per\square\cm})}
 & \multicolumn{2}{c}{(kpc)}
 & \colhead{}
 & \colhead{(\si{\per\cubic\cm})}
 & \colhead{}
}
\startdata
\multicolumn{11}{c}{Reddened Stars} \\ \hline 
BD+35d4258 & 21.26 & 0.05 & 21.24 & 0.05 & 19.55 & 0.07 & 1.89 & 0.19 & 0.04 & 0.31 & 6 \\
BD+53d2820 & 21.39 & 0.05 & 21.35 & 0.05 & 20.08 & 0.17 & 3.18 & 0.41 & 0.10 & 0.25 & 6 \\
BD+56d524 & 21.57 & 0.16 & 21.38 & 0.16 & 20.82 & 0.28 & 2.19 & 0.21 & 0.36 & 0.55 & 1 \\
HD001383 & 21.54 & 0.04 & 21.46 & 0.04 & 20.44 & 0.13 & 3.04 & 0.30 & 0.16 & 0.37 & 6 \\
HD013268\tablenotemark{a} & 21.43 & 0.04 & 21.34 & 0.05 & 20.42 & 0.08 & 2.80 & 0.28 & 0.19 & 0.31 & 6 \\
HD014250 & 21.63 & 0.13 & 21.52 & 0.15 & 20.66 & 0.18 & 1.20 & 0.11 & 0.22 & 1.14 & 4 \\
HD014434\tablenotemark{a} & 21.48 & 0.08 & 21.37 & 0.09 & 20.54 & 0.14 & 2.98 & 0.30 & 0.23 & 0.33 & 2 \\
HD015558 & 21.66 & 0.14 & 21.52 & 0.18 & 20.80 & 0.07 & 2.02 & 0.36 & 0.28 & 0.73 & 3 \\
HD017505 & 21.48 & 0.10 & 21.26 & 0.15 & 20.79 & 0.07 & 2.56 & 0.22 & 0.40 & 0.39 & 3 \\
HD023060 & 21.51 & 0.06 & 21.40 & 0.07 & 20.55 & 0.07 & 0.53 & 0.01 & 0.22 & 1.96 & 1 \\
HD027778 & 21.36 & 0.08 & 21.10 & 0.12 & 20.72 & 0.08 & 0.22 & 0.00 & 0.45 & 3.36 & 2 \\
HD037903 & 21.46 & 0.06 & 21.16 & 0.09 & 20.85 & 0.07 & 0.40 & 0.01 & 0.49 & 2.34 & 2 \\
HD038087 & 21.59 & 0.12 & 21.48 & 0.15 & 20.65 & 0.06 & 0.34 & 0.01 & 0.23 & 3.78 & 4 \\
HD045314\tablenotemark{a} & 21.26 & 0.06 & 21.05 & 0.09 & 20.54 & 0.06 & 0.80 & 0.08 & 0.38 & 0.74 & 1 \\
HD046056 & 21.53 & 0.10 & 21.38 & 0.14 & 20.71 & 0.07 & 1.46 & 0.13 & 0.30 & 0.76 & 3 \\
HD046150\tablenotemark{a} & 21.40 & 0.09 & 21.26 & 0.12 & 20.54 & 0.08 & 1.47 & 0.15 & 0.28 & 0.55 & 3 \\
HD046202 & 21.52 & 0.10 & 21.39 & 0.13 & 20.63 & 0.07 & 1.30 & 0.25 & 0.26 & 0.83 & 1 \\
HD047129 & 21.36 & 0.07 & 21.26 & 0.09 & 20.38 & 0.07 & 1.45 & 0.20 & 0.21 & 0.51 & 1 \\
HD047240 & 21.31 & 0.10 & 21.20 & 0.12 & 20.35 & 0.09 & 2.01 & 0.30 & 0.22 & 0.33 & 3 \\
HD047417\tablenotemark{a} & 21.28 & 0.07 & 21.13 & 0.09 & 20.45 & 0.09 & 1.23 & 0.12 & 0.29 & 0.50 & 3 \\
HD062542 & 21.69 & 0.11 & 21.54 & 0.12 & 20.85 & 0.21 & 0.39 & 0.00 & 0.29 & 4.11 & 1 \\
HD073882\tablenotemark{a} & 21.59 & 0.07 & 21.11 & 0.15 & 21.11 & 0.07 & 0.83 & 0.08 & 0.67 & 1.51 & 4 \\
HD091651\tablenotemark{a} & 21.16 & 0.06 & 21.15 & 0.06 & 19.08 & 0.11 & 3.98 & 0.40 & 0.02 & 0.12 & 3 \\
HD093222\tablenotemark{a} & 21.49 & 0.03 & 21.47 & 0.03 & 19.77 & 0.09 & 2.47 & 0.25 & 0.04 & 0.40 & 6 \\
HD093250\tablenotemark{a} & 21.46 & 0.14 & 21.39 & 0.15 & 20.32 & 0.14 & 2.20 & 0.22 & 0.15 & 0.42 & 3 \\
HD093827 & 21.41 & 0.18 & 21.34 & 0.19 & 20.25 & 0.10 & 3.60 & 0.47 & 0.14 & 0.23 & 1 \\
HD096675 & 21.88 & 0.09 & 21.80 & 0.10 & 20.82 & 0.11 & 0.16 & 0.00 & 0.17 & 15.28 & 1 \\
HD096715\tablenotemark{a} & 21.39 & 0.11 & 21.20 & 0.16 & 20.65 & 0.06 & 3.15 & 0.32 & 0.36 & 0.25 & 3 \\
HD099872\tablenotemark{b} & 21.47 & 0.07 & 21.35 & 0.08 & 20.55 & 0.11 & \multicolumn{2}{c}{\nodata} & 0.24 & \nodata & 1 \\
HD099890\tablenotemark{a} & 21.14 & 0.05 & 21.12 & 0.05 & 19.53 & 0.09 & 3.22 & 0.32 & 0.05 & 0.14 & 6 \\
HD100213\tablenotemark{a} & 21.32 & 0.06 & 21.18 & 0.07 & 20.45 & 0.09 & 2.15 & 0.21 & 0.27 & 0.31 & 3 \\
HD101190\tablenotemark{a} & 21.36 & 0.03 & 21.24 & 0.04 & 20.44 & 0.07 & 2.09 & 0.21 & 0.24 & 0.35 & 6 \\
HD101205\tablenotemark{a} & 21.29 & 0.06 & 21.20 & 0.07 & 20.26 & 0.15 & 1.64 & 0.16 & 0.19 & 0.39 & 3 \\
HD103779\tablenotemark{a} & 21.21 & 0.03 & 21.17 & 0.03 & 19.86 & 0.11 & 3.99 & 0.40 & 0.09 & 0.13 & 6 \\
HD122879 & 21.39 & 0.04 & 21.31 & 0.04 & 20.31 & 0.08 & 2.23 & 0.19 & 0.17 & 0.36 & 6 \\
HD124979\tablenotemark{a} & 21.38 & 0.05 & 21.27 & 0.06 & 20.44 & 0.07 & 3.09 & 0.31 & 0.23 & 0.25 & 6 \\
HD147888 & 21.73 & 0.12 & 21.68 & 0.13 & 20.47 & 0.08 & 0.09 & 0.00 & 0.11 & 19.04 & 6 \\
HD148422\tablenotemark{a} & 21.30 & 0.05 & 21.24 & 0.06 & 20.13 & 0.07 & 8.26 & 0.83 & 0.13 & 0.08 & 6 \\
HD149404 & 21.58 & 0.10 & 21.40 & 0.14 & 20.80 & 0.07 & 1.27 & 0.28 & 0.33 & 0.97 & 3 \\
HD151805 & 21.41 & 0.03 & 21.33 & 0.03 & 20.36 & 0.07 & 1.59 & 0.14 & 0.18 & 0.53 & 6 \\
HD152233\tablenotemark{a} & 21.37 & 0.09 & 21.29 & 0.10 & 20.31 & 0.11 & 1.52 & 0.15 & 0.17 & 0.50 & 3 \\
HD152234 & 21.43 & 0.10 & 21.32 & 0.12 & 20.50 & 0.08 & 1.53 & -7.01 & 0.23 & 0.58 & 3 \\
HD152248\tablenotemark{a} & 21.66 & 0.14 & 21.62 & 0.14 & 20.26 & 0.07 & 1.61 & 0.16 & 0.08 & 0.91 & 1 \\
HD152249 & 21.45 & 0.06 & 21.38 & 0.05 & 20.34 & 0.25 & 1.95 & 0.21 & 0.15 & 0.47 & 6 \\
HD152723\tablenotemark{a} & 21.49 & 0.12 & 21.43 & 0.13 & 20.30 & 0.09 & 1.94 & 0.19 & 0.13 & 0.52 & 3 \\
HD157857\tablenotemark{a} & 21.43 & 0.07 & 21.26 & 0.09 & 20.65 & 0.07 & 2.56 & 0.26 & 0.33 & 0.34 & 2 \\
HD160993 & 21.20 & 0.10 & 21.18 & 0.10 & 19.49 & 0.04 & 3.11 & 0.51 & 0.04 & 0.16 & 3 \\
HD163522 & 21.16 & 0.05 & 21.14 & 0.05 & 19.59 & 0.14 & 3.87 & 1.10 & 0.05 & 0.12 & 6 \\
HD164816\tablenotemark{a} & 21.22 & 0.12 & 21.18 & 0.13 & 19.89 & 0.05 & 1.08 & 0.11 & 0.09 & 0.50 & 3 \\
HD164906 & 21.29 & 0.08 & 21.20 & 0.09 & 20.24 & 0.07 & 1.19 & 0.08 & 0.18 & 0.53 & 3 \\
HD165052\tablenotemark{a} & 21.41 & 0.09 & 21.36 & 0.10 & 20.16 & 0.11 & 1.37 & 0.14 & 0.11 & 0.61 & 3 \\
HD167402\tablenotemark{a} & 21.22 & 0.03 & 21.13 & 0.03 & 20.17 & 0.09 & 7.61 & 0.76 & 0.18 & 0.07 & 6 \\
HD167771\tablenotemark{a} & 21.33 & 0.08 & 21.08 & 0.12 & 20.68 & 0.10 & 1.40 & 0.14 & 0.44 & 0.50 & 3 \\
HD168076\tablenotemark{a} & 21.73 & 0.21 & 21.65 & 0.23 & 20.68 & 0.07 & 1.97 & 0.20 & 0.18 & 0.89 & 3 \\
HD168941\tablenotemark{a} & 21.26 & 0.04 & 21.18 & 0.05 & 20.21 & 0.07 & 3.72 & 0.37 & 0.18 & 0.16 & 6 \\
HD178487\tablenotemark{a} & 21.36 & 0.06 & 21.22 & 0.07 & 20.50 & 0.09 & 5.22 & 0.52 & 0.28 & 0.14 & 6 \\
HD179406\tablenotemark{a} & 21.55 & 0.08 & 21.42 & 0.10 & 20.67 & 0.06 & 0.21 & 0.02 & 0.26 & 5.50 & 1 \\
HD179407\tablenotemark{a} & 21.30 & 0.06 & 21.20 & 0.07 & 20.31 & 0.07 & 7.72 & 0.77 & 0.20 & 0.08 & 6 \\
HD185418\tablenotemark{a} & 21.41 & 0.03 & 21.19 & 0.03 & 20.70 & 0.07 & 0.78 & 0.08 & 0.39 & 1.06 & 6 \\
HD188001 & 21.14 & 0.07 & 21.03 & 0.08 & 20.18 & 0.08 & 1.77 & 0.15 & 0.22 & 0.25 & 3 \\
HD190603 & 21.72 & 0.10 & 21.63 & 0.12 & 20.68 & 0.07 & 2.71 & 0.73 & 0.18 & 0.62 & 1 \\
HD192639\tablenotemark{a} & 21.47 & 0.06 & 21.29 & 0.09 & 20.70 & 0.07 & 2.14 & 0.21 & 0.34 & 0.45 & 2 \\
HD197770 & 21.42 & 0.12 & 21.01 & 0.23 & 20.90 & 0.12 & 0.87 & 0.02 & 0.61 & 0.97 & 1 \\
HD198781 & 21.16 & 0.04 & 20.93 & 0.05 & 20.48 & 0.07 & 0.91 & 0.04 & 0.42 & 0.52 & 6 \\
HD199579\tablenotemark{a} & 21.25 & 0.08 & 21.04 & 0.11 & 20.53 & 0.09 & 0.92 & 0.09 & 0.38 & 0.62 & 3 \\
HD200775 & 21.53 & 0.07 & 21.10 & 0.10 & 21.03 & 0.10 & 0.36 & 0.01 & 0.63 & 3.09 & 5 \\
HD203938 & 21.70 & 0.10 & 21.48 & 0.15 & 21.01 & 0.07 & 0.22 & 0.04 & 0.40 & 7.42 & 4 \\
HD206267\tablenotemark{a} & 21.47 & 0.03 & 21.22 & 0.04 & 20.81 & 0.05 & 0.73 & 0.07 & 0.44 & 1.31 & 6 \\
HD206773\tablenotemark{a} & 21.24 & 0.04 & 21.09 & 0.05 & 20.42 & 0.09 & 0.69 & 0.07 & 0.30 & 0.82 & 6 \\
HD207198\tablenotemark{a} & 21.51 & 0.05 & 21.28 & 0.06 & 20.83 & 0.07 & 1.04 & 0.10 & 0.42 & 1.02 & 6 \\
HD209339\tablenotemark{a} & 21.25 & 0.09 & 21.16 & 0.11 & 20.24 & 0.08 & 1.08 & 0.11 & 0.19 & 0.54 & 1 \\
HD216898\tablenotemark{a} & 21.82 & 0.19 & 21.66 & 0.25 & 21.02 & 0.07 & 0.91 & 0.09 & 0.31 & 2.37 & 1 \\
HD239729 & 21.58 & 0.10 & 21.15 & 0.15 & 21.07 & 0.12 & 0.93 & 0.03 & 0.62 & 1.31 & 4 \\
HD326329 & 21.56 & 0.12 & 21.51 & 0.13 & 20.27 & 0.04 & 1.40 & 0.10 & 0.10 & 0.84 & 1 \\
HD332407\tablenotemark{a} & 21.35 & 0.11 & 21.24 & 0.14 & 20.40 & 0.07 & 2.56 & 0.26 & 0.22 & 0.28 & 3 \\
\multicolumn{11}{c}{Comparison Stars} \\ \hline 
BD+32d270 & 20.79 & 0.09 & 20.78 & 0.09 & 18.69 & 0.04 & 1.89 & 0.19 & 0.02 & 0.11 & 3 \\
BD+52d3210 & 21.47 & 0.11 & 21.45 & 0.11 & 19.88 & 0.07 & 3.18 & 0.41 & 0.05 & 0.30 & 1 \\
HD037332 & 21.58 & 0.15 & 21.58 & 0.14 & 15.69 & 0.19 & 2.19 & 0.21 & 0.00 & 0.56 & 1 \\
HD037525 & 21.44 & 0.12 & 21.44 & 0.12 & 16.24 & 1.50 & 3.04 & 0.30 & 0.00 & 0.29 & 1 \\
HD051013 & 21.08 & 0.14 & 21.08 & 0.13 & 14.40 & 0.09 & 1.61 & 0.12 & 0.00 & 0.24 & 1 \\
HD075309 & 21.18 & 0.04 & 21.10 & 0.04 & 20.11 & 0.07 & 1.20 & 0.11 & 0.17 & 0.41 & 6 \\
HD091824\tablenotemark{a} & 21.16 & 0.03 & 21.12 & 0.03 & 19.75 & 0.07 & 2.75 & 0.28 & 0.08 & 0.17 & 6 \\
HD091983 & 21.23 & 0.04 & 21.15 & 0.04 & 20.16 & 0.09 & 2.02 & 0.36 & 0.17 & 0.27 & 6 \\
HD093028\tablenotemark{a} & 21.17 & 0.13 & 21.15 & 0.13 & 19.50 & 0.09 & 2.97 & 0.30 & 0.04 & 0.16 & 1 \\
HD094493 & 21.18 & 0.09 & 21.10 & 0.11 & 20.13 & 0.08 & 0.53 & 0.01 & 0.18 & 0.93 & 1 \\
HD097471\tablenotemark{a} & 21.39 & 0.11 & 21.36 & 0.11 & 19.89 & 0.07 & 2.78 & 0.28 & 0.06 & 0.29 & 1 \\
HD100276\tablenotemark{a} & 21.23 & 0.08 & 21.19 & 0.09 & 19.83 & 0.06 & 2.94 & 0.29 & 0.08 & 0.19 & 3 \\
HD104705\tablenotemark{a} & 21.20 & 0.04 & 21.15 & 0.04 & 19.95 & 0.10 & 4.18 & 0.42 & 0.11 & 0.12 & 6 \\
HD114444 & 21.24 & 0.09 & 21.20 & 0.09 & 19.83 & 0.11 & 0.81 & 0.03 & 0.08 & 0.69 & 3 \\
HD116852\tablenotemark{a} & 21.02 & 0.03 & 20.96 & 0.03 & 19.80 & 0.07 & 4.88 & 0.49 & 0.12 & 0.07 & 6 \\
HD172140\tablenotemark{a} & 21.12 & 0.08 & 21.11 & 0.08 & 19.21 & 0.13 & 6.54 & 0.65 & 0.02 & 0.07 & 3 \\
HD210809\tablenotemark{a} & 21.34 & 0.04 & 21.31 & 0.04 & 19.92 & 0.17 & 3.88 & 0.39 & 0.08 & 0.18 & 6 \\
HD235874 & 21.46 & 0.11 & 21.45 & 0.11 & 19.64 & 0.08 & 1.45 & 0.20 & 0.03 & 0.65 & 1 \\
\enddata
\tablenotetext{a}{Distance taken from \citet{2021ApJ...911...55S}.}
\tablenotetext{b}{No Gaia parallax or photometric distance provided.}
\tablecomments{The sources for the \nhi\ data are numbered in the rightmost column, and given above. Additional quantities shown: 
Distance $d$, either photometric distance values from \citet{2021ApJ...911...55S}, or values based on Gaia DR2 parallax.
Derived molecular fraction \fhtwo.
Log of density estimate $\nh / d$.}
\tablerefs{
    (1) This work
    (2) \citet{2006ApJ...641..327C};
    (3) \citet{1994ApJ...427..274D};
    (4) \citet{1990ApJS...72..163F};
    (5) \citet{1994ApJ...430..630B};
    (6) \citet{2021ApJ...911...55S};}
\end{deluxetable*}